
\input harvmac
\def\npb#1(#2)#3{{ Nucl. Phys. }{B#1} (#2) #3}
\def\plb#1(#2)#3{{ Phys. Lett. }{#1B} (#2) #3}
\def\pla#1(#2)#3{{ Phys. Lett. }{#1A} (#2) #3}
\def\prl#1(#2)#3{{ Phys. Rev. Lett. }{#1} (#2) #3}
\def\mpla#1(#2)#3{{ Mod. Phys. Lett. }{A#1} (#2) #3}
\def\ijmpa#1(#2)#3{{ Int. J. Mod. Phys. }{A#1} (#2) #3}
\def\cmp#1(#2)#3{{ Commun. Math. Phys. }{#1} (#2) #3}
\def\cqg#1(#2)#3{{ Class. Quantum Grav. }{#1} (#2) #3}
\def\jmp#1(#2)#3{{ J. Math. Phys. }{#1} (#2) #3}
\def\anp#1(#2)#3{{ Ann. Phys. }{#1} (#2) #3}
\def\prd#1(#2)#3{{ Phys. Rev.} {D\bf{#1}} (#2) #3}

\def\inbar{\,\vrule height1.5ex width.4pt depth0pt}
\def\IQ{\relax\,\hbox{$\inbar\kern-.3em{\rm Q}$}}
\def\IB{\relax{\rm I\kern-.18em B}}
\def\IC{\relax\hbox{$\inbar\kern-.3em{\rm C}$}}
\def\IP{\relax{\rm I\kern-.18em P}}
\def\IR{\relax{\rm I\kern-.18em R}}
\def\ZZ{\relax\ifmmode\mathchoice
{\hbox{Z\kern-.4em Z}}{\hbox{Z\kern-.4em Z}}
{\lower.9pt\hbox{Z\kern-.4em Z}}
{\lower1.2pt\hbox{Z\kern-.4em Z}}\else{Z\kern-.4em Z}\fi}

\Title{}{Arithmetic Properties of Mirror Map and Quantum Coupling
\footnote{${}^\diamond$}{Research supported by grant DE-FG02-88-ER-25065.}}

\centerline{
Bong H. Lian
and Shing-Tung Yau\footnote{ }{emails:  lian@math.harvard.edu,
yau@math.harvard.edu}}
\bigskip
\centerline{
            \it Department of Mathematics}
\centerline{\it Harvard University}
\centerline{\it Cambridge, MA 02138, USA}
\vskip .2in

Abstract:  We study some arithmetic properties of the mirror maps and the
quantum Yukawa coupling for some 1-parameter deformations of Calabi-Yau
manifolds.  First we use the Schwarzian differential equation, which we derived
previously, to characterize the mirror map in each case. For algebraic K3
surfaces, we solve the equation in terms of the $J$-function. By deriving
explicit modular relations we prove that some K3 mirror maps are algebraic over
the genus zero function field ${\bf Q}(J)$. This leads to a uniform proof that
those mirror maps have integral Fourier coefficients. Regarding the maps as
Riemann mappings, we prove that they are genus zero functions. By virtue  of
the Conway-Norton conjecture (proved by Borcherds using
Frenkel-Lepowsky-Meurman's
Moonshine module), we find that these maps are actually the reciprocals of the
Thompson series for certain conjugacy classes in the Griess-Fischer group. This
also gives, as an immediate consequence, a second proof that those mirror maps
are integral. We thus conjecture a surprising connection between K3 mirror maps
and the Thompson series. For threefolds, we construct a formal nonlinear ODE
for the quantum coupling reduced $mod\ p$. Under the mirror hypothesis and an
integrality assumption, we derive $mod~p$ congruences for the Fourier
coefficients. For the quintics, we deduce (at least for $5\not{|}d$) that
the degree $d$ instanton numbers $n_d$ are divisible by $5^3$ -- a fact
first conjectured by Clemens.

\Date{Revised 12/94} 


\newsec{Introduction}

For background on Mirror Symmetry, the readers are referred to the reference
\ref\cdgp{P. Candelas, X. De la Ossa, P. Green and L. Parkes, Nucl. Phys. B359
(1991) 21.}
\ref\emm{Essays on Mirror Manifolds, Ed. S.-T. Yau,
(International Press, Hong Kong 1992)} (see especially the articles therein by
Greene-Plesser, Candelas-de la Ossa-Green-Parkes, Katz, Morrison, Vafa and
Witten.)

It is known that the so-called mirror map and the quantum coupling have many
interesting number theoretic properties based on numerical experiments -- as
previously observed by many \cdgp\ref\bvs{V. Batyrev and D. van Straten, {\sl
Generalized hypergeometric functions and rational curves on Calabi-Yau complete
intersections in toric varieties}, preprint 1993.}\ref\hktyI{S. Hosono, A.
Klemm, S. Theisen and S.-T. Yau:
           {\sl Mirror symmetry, mirror map and applications
           to Calabi-Yau hypersurfaces},
           HUTMP-93/0801, LMU-TPW-93-22  (hep-th/9308122), to be
           published in Commun. Math. Phys.}\ref\hktyII{S. Hosono, A. Klemm,
           S.Theisen and S. T. Yau, {\sl Mirror symmetry, mirror map
           and applications to complete intersection Calabi-Yau
           spaces\/}, Preprint HUTMP-94-02, hep-th 9406055}. For example the
Fourier coefficients of the mirror map appears to be integral in
all known cases. In some cases, the coefficients even appear to be
alternating. The instanton numbers $n_d$ in the quantum coupling on the
other hand, apparently have some striking divisibility property. Clemens
conjectured that for the quintics in ${\bf CP}^4$, $5^3|n_d$ for all $d$
\ref\clemens{H. Clemens,Proc. ICM, Berkeley (1986) 634.}\footnote{*}{We thank
S. Katz for pointing out some references to us.}. This was supported by Katz'
proof that $5|n_d$, along with similar divisibility properties for other
manifolds \ref\skatz{S. Katz, Math. Z. 191 (1986) 293.}. Our main motivation
here is to develop some techniques, along with mirror symmetry, to understand
some of these remarkable ``arithmetic'' properties.

The technique for studying the mirror map is based on the following simple
idea: fix a known integral series $f(q)$. Study when is $z(q)$ commensurable
with $f(q)$ (ie. when do $z(q),f(q)$ satisfy a polynomial relation)? Hopefully
when enough is known about $f(q)$,  then given a polynomial $p(X,Y)$ we can
understand some of the arithmetic properties of the root $z(q)$ to
$p(f(q),Y)=0$. Note that this is still a difficult Diophantine type problem in
general, involving infinitely many variables consisting of the Fourier
coefficients of $z(q)$.

However, a lot is known about the j-function both number theoretically and
geometrically. Thus it is natural to try to find commensurability relations
(also known as modular relations) between $j(q),z(q)$. We will see that this
idea works well in the case of elliptic curves and K3 surfaces.

The technique for studying the instanton numbers is based on the fact that
there is a canonical polynomial ODE for the quantum coupling, which is defined
over ${\bf Q}$. This raises the possibility of deriving similar equations, but
reduced $mod~p$. The $mod~p$ arithmetic properties of the quantum couplings
should then be reflected in these reduced equations.
In \ref\klryA{A. Klemm, B.H. Lian, S.S. Roan and S.T. Yau, A Note on ODEs from
Mirror Symmetry, To appear in the Conference Proceedings in honor of I.
Gel'fand}\ref\klryB{A. Klemm, B.H. Lian, S.S. Roan and S.T. Yau, Differential
Equations from Mirror Symmetry, in preparation.}, in collaboration with A.
Klemm and S.S. Roan we have constructed an ODE over ${\bf Q}$. However, its
$mod~p$ counterpart derived here appears much simpler and more manageable. We
now summarize our discussion.

First we discuss, along the lines of \hktyI\hktyII, the construction of
deformation coordinates based on which the mirror map is defined. Then
We review the polynomial differential equations for the mirror maps, studied in
\klryA\klryB. First we classify the analytic solutions, on a disk, to the
$n^{th}$ (n=2,3 or 4) Schwarzian equations associated to certain $n^{th}$ order
linear ODEs of Fuchsian type. As a consequence, the so-called mirror map $z$
can be given a simple characterization. Corresponding to n=2,3 or 4, there is a
universal family of polynomials whose evaluation at certain integral points
recovers the Fourier coefficients of a mirror map.

We consider some examples in which the linear ODEs are the Picard-Fuchs
equations of several distinguished families of smooth Calabi-Yau varieties in
weighted projective spaces. We revisit the case of elliptic curves (n=2).

In the case of n=3, we consider some distinguished families of K3 surfaces in
weighted projective spaces. We prove that the the mirror map $z$ for each of
those families is algebraic over the function field ${\bf Q}(J)$ generated by
the Dedekind-Klein J-function -- a rather surprising connection between modular
functions and K3 surfaces. Our result on the n=3 Schwarzian equation is an
important tool in this connection. We then use our explicit modular relations
to give a uniform proof that the Fourier coefficients of $z$ are integral -- a
fact which has been previously observed experimentally. This is also the first
confirmation of the integrality property in the case of K3 surfaces.

We then discuss a mysterious connection between our mirror maps and the
Thompson series. We offer some speculation as to why the two might be related.
We conjecture that any 1-parameter deformation of algebraic K3, determined by
an orbifold construction, gives rise to a Thompson series.

In the last section under an integrality assumption and the Mirror Hypothesis,
we study our differential equations reduced mod $p$. We derived some general
$mod~p$ congruences and then specialize to $p=2,3,5,7$. For the quintic
hypersurface, we deduce Clemens' conjecture that the degree $d$ instanton
numbers $n_d$ are divisible by $5^3$ (at least for $5\not{|}d$).

{\bf Acknowledgements}: B.H.L. thanks W. Feit, A. Klemm, T. Tamagawa, S.
Theisen, A. Todorov and G. Zuckerman for helpful discussions.

\newsec{Deformation coordinates}
\seclab\defcoord

The purpose here is to review the orbifold construction of the mirror map and
to give an explicit description of the deformation coordinates for complex
structures. The orbifold construction in weighted projective spaces has now
been superceded by toric geometry construction \ref\batyrev{V. Batyrev,
{\sl Dual Polyhedra and the Mirror Symmetry
      for Calabi-Yau Hypersurfaces in Toric Varieties}, Univ. Essen
      Preprint (1992), to appear in  Journal of Alg. Geom.}\hktyI. But in the
former, the description of the deformation coordinates can be made rather
explicit. Here we will adopt the multi-indexed convention for monomials
$y^I=y_1^{i_1}\cdots y_m^{i_m}$ whenever the meaning of the variables $y$ is
clear.

Let $X$ be a $l:=(n-k-1)$-dimensional Calabi-Yau variety defined as the zero
locus of $k$ homogeneous polynomials $p_1,..,p_k$ whose degrees are
$d_1,..,d_k$, in the weighted projective space ${\bf P}^{n-1}[w]$. Here
$w=(w_1,..,w_n)$ are the weights consisting of coprime positive integers.
Suppose that $X$ has
 a mirror $X^*$ given by an orbifold construction \ref\roan{S.-S. Roan, Intern.
J. Math. 1 (1990) 211-232.}\ref\gp{B. Greene and R. Plesser, Nucl. Phys. B 338
(1990),15-37.}\cdgp. (Thus $X^*$ is assumed to have the usual mirror Hodge
diamond.) The space $X^*$ is a resolution $\widehat{X/G}$ of the singular
quotient $X/G$, where $G$ is a finite abelian group acting on the homogeneous
coordinates $x_i$
of the ambient space ${\bf P}^{n-1}[w]$ by characters $\chi_i:G\rightarrow
S^1$. They are assumed to satisfy $\chi_1\cdots\chi_n=1$, which ensures that
the holomorphic top form on $X$ is $G$-invariant, and hence induces a form on
$X^*$. We now want to describe some complex structure deformations of $X^*$.
We do so by
describing a family of $G$-invariant holomorphic top forms $\Omega(z)$ on $X$.
Thus ultimately, our
deformation space $M$ will be parameterizing a family of $G$-invariant
deformations of the $p_j$.

 Fix $m_j$ $G$-invariants monomials which we denote by $x^{I_i}$, $i\in
N_j:=\{(m_1+..+m_{j-1})+1,..,(m_1+..+m_{j-1})+m_j\}$. Let $m:=m_1+\cdots+m_k$.
Let $F$ be the following holomorphic family, fibered over $({\bf C}^\times)^m$,
of varieties. Its fiber at $a=(a_1,..,a_m)$ is defined as the zero locus of
\eqn\dumb{\tilde{p}_j(a;x)=\sum_{i\in N_j} a_i x^{I_i}.}
We assume that $X$ is the fiber at some limiting value of $a$ and that
the generic fibers are homeomorphic to $X$. Now two fibers can be biholomorphic
simply by coordinate transformation. Specifically for any
$\alpha=(\alpha_1,..,\alpha_{n+k})\in ({\bf C}^\times)^{n+k}$, the
transformation $x_i\mapsto x_i/\alpha_i$,
$a_i\mapsto a_i\alpha_1^{i_1}\cdots\alpha_n^{i_n}\alpha_{n+j} \equiv
a_i\alpha^{J_i}$ where $(i_1,..,i_n)=I_i$, transforms $\tilde{p}_j\mapsto
\alpha_{n+j}\tilde{p}_j$. Call this transformation group $H$, and consider the
quotient family $F/H\rightarrow ({\bf C}^\times)^m/H=M$. The base $M$ will be
our deformation space. The isotropy group of the $H$-action on $({\bf
C}^\times)^m$ is $H'=\{\alpha|\alpha^{J_i}=1,~all~ i\}$. The conditions
$\alpha^{J_i}=1$ mean that the dot products $J_i\cdot(log\ \alpha_1,..,log\
\alpha_{n+k})=0$. Or if $K$ is the $m\times (n+k)$ {\it matrix of exponents}
whose rows are the $J_i$, then $K\cdot(log\ \alpha)^t=0$. Thus $rk\ K=dim\
H-dim\ H'$, implying that $dim\ M=m-rk\ K$.

We wish to construct some coordinates on $M$. Now $H$ acts on the coordinate
ring ${\bf C}[a_1^{\pm1},..,a_m^{\pm1}]$ of $({\bf C}^\times)^m$ by $a_i\mapsto
a_i \alpha^{J_i}$. A simple way to coordinatize $M=({\bf C}^\times)^m/H$ would
be to find enough suitable $H$-invariant functions $f:({\bf
C}^\times)^m\rightarrow {\bf C}^\times$.  An $H$-invariant Laurent monomial
$a^\mu$ is one with $\mu_1J_1+\cdots +\mu_mJ_m=0$. The set of such $\mu$ is the
lattice $L=(ker\ K^t)\cap{\bf Z}^m$ which has rank $dim(ker\ K^t)=m-rk\ K^t
=dim\ M$. The subalgebra of $H$-invariants in ${\bf
C}[a_1^{\pm1},..,a_m^{\pm1}]$ is the group algebra ${\bf C}[L]$ which is
canonically generated by $a^{\pm B}$, for a given basis $\{B\}$ of $L$. Thus
for every basis $\{B\}$, we get a canonical set of functions $z=(a^B)$ globally
defined on $M$. The $z$ will be our deformation coordinates.

Since any two bases $\{B\},\{B'\}$ are related by $B=\sum m_{BB'}B'$ where
$(m_{BB'})$ is an integral matrix, $z=(a^B)\mapsto z'=(a^{B'})$ is a birational
change. Note that if the $z$ take the value of {\it integral} q-series, then so
do the image $z'$ under the above change. Thus when the $z_i$ takes the value
of the mirror map which has a q-series expansion, the question of integrality
of the Fourier coefficients is independent of the choice of basis of $L$. In
case $dim\ M=1$, which is all we are going to deal with here, the coordinate
$z$ above is unique up to $z\mapsto 1/z$. But demanding that $z=0$ is the point
with maximum unipotent monodromy \ref\morrison{D. R. Morrison,  in Essays
on mirror manifolds Ed. S.-T.Yau (International Press Singapore 1992)} for the
Picard-Fuchs equation, fixes the choice completely.

\subsec{Some examples}

Let's first consider the simplest example: cubics in ${\bf P}^2$. Let $G$ be a
cyclic group of order 3 and act on the homogeneous coordinates by $x_i\mapsto
\xi^i x_i$ where $\xi=e^{2\pi i/3}$. Then the $G$-invariant cubic monomials are
$x_1^3,x_2^3,x_3^3,x_1x_2x_3$. We consider the invariant family $F$ which is
the zero locus of the polynomial
\eqn\dumb{\tilde{p}(a;x)=a_1x_1^3+a_2x_2^3+a_3x_3^3+a_4x_1x_2x_3.}
The group $H=({\bf C}^\times)^{3+1}$ acts on $F$ as described above. On the
base space $({\bf C}^\times)^4$, it acts by $\alpha: a\mapsto (\alpha_1^3
a_1,\alpha_2^3 a_2,\alpha_3^3 a_3, \alpha_1\alpha_2\alpha_3 a_4)\alpha_4$. We
see that the $H$-invariant functions are generated by $z=a_1 a_2 a_3 a_4^{-3}$.
It's easy to check that this function defines an isomorphism $({\bf
C}^\times)^4/H\rightarrow {\bf C}^\times$.

Consider now the sextics in ${\bf P}^3[1,1,2,2]$. Let $G$ be a finite abelian
group of type $(6,3,3)$, which acts on the homogeneous coordinates by
\eqn\dumb{x_1\mapsto\xi_1\xi_2\xi_3 x_1,~
x_2\mapsto \xi_1^{-1}x_2,~
x_3\mapsto \xi_2^{-1}x_3,~
x_4\mapsto \xi_3^{-1}x_4}
where $\xi_1,\xi_2,\xi_3$ are respectively arbitrary 6th, 3rd, 3rd roots of 1.
The $G$-invariants monomials are
$x_1^6,x_2^6,x_3^3,x_4^3,x_1x_2x_3x_4,x_1^3x_2^3$.
As before we can write down a linear sum of them with coefficients
$a_1,..,a_6.$
The matrix of exponents $K$ defined above is the $6\times5$ matrix with rows:
$(6,0,0,0,1),(0,6,0,0,1),(0,0,3,0,1),(0,0,0,3,1),(1,1,1,1,1),(3,3,0,0,1)$. The
lattice $L$ therefore has a base $(1,1,0,0,0,-2),(0,0,1,1,-3,1)$. This gives us
the deformation coordinates $z=(z_1,z_2)$ with $z_1=a_1 a_2 a_6^{-2}$, $z_2=a_3
a_4 a_5^{-3} a_6$.

As a third example, we consider the sextics in ${\bf P}^3[1,1,1,3]$. Let $G$ be
a finite abelian group of type $(6,6,2)$, which acts on the homogeneous
coordinates by
\eqn\dumb{x_1\mapsto\xi_1\xi_2\xi_3 x_1,~
x_2\mapsto \xi_1^{-1}x_2,~
x_3\mapsto \xi_2^{-1}x_3,~
x_4\mapsto \xi_3^{-1}x_4}
where $\xi_1,\xi_2,\xi_3$ are respectively arbitrary 6th, 6th, 2nd roots of 1.
The $G$-invariants monomials are
$x_1^6,x_2^6,x_3^6,x_4^2,x_1x_2x_3x_4,x_1^2x_2^2x_3^2$. As before we can write
down a general linear sum of them to define our family of sextics. But note
that any sum of the form $ax_4^2+bx_1x_2x_3x_4+cx_1^2x_2^2x_3^2$ can be written
as
$ax_4^2+b'x_1x_2x_3x_4$, by a suitable redefinition $x_1\mapsto x_1,x_2\mapsto
x_2, x_3\mapsto x_3$, $x_4\mapsto x_4+\lambda x_1x_2x_3$. Thus we consider the
family:
\eqn\dumb{p(a;x)=a_1x_1^6+a_2x_2^6+a_3x_3^6+a_4x_4^2+a_5x_1x_2x_3x_4.}
The matrix of exponents $K$ is the $5\times5$ matrix with rows: $(6,0,0,0,1)$,
$(0,6,0,0,1)$, $(0,0,3,0,1)$, $(0,0,0,3,1)$, $(1,1,1,1,1)$. The lattice $L$
therefore has a base $(1,1,1,3,-6)$. This gives us the deformation coordinate
$z=a_1 a_2 a_3 a_4^3a_5^{-6}$.

\subsec{Definition of the mirror map}

Following \cdgp\morrison\hktyI, we now define the mirror map in terms of the
coordinates $z$ by studying variation of the periods for $X^*$. By
construction, the holomorphic top form $\Omega(z)$ on $X$ induces a form on
$X^*$. Integrating the form against the $G$-invariant cycles, we obtain periods
for $X^*$. By the Dwork-Griffith-Katz reduction method, we get a system of
Picard-Fuchs differential equation(s) for those periods. If $z=0$ is a maximal
unipotent point, then the system admits a unique solution $\omega_0(z)$ which
is holomorphic near $z=0$ with $\omega_0(0)=1$, and $s:=dim~M$ independent
solutions of the form $\omega_i(z)=\omega_0(z)log~z_i +g_i(z)$ where the $g_i$
are holomorphic near $z=0$ with $g_i(0)=0$. We call the mapping defined by
$q_j:=e^{2\pi i t_j}=z_j e^{{g_j(z)\over\omega_0(z)}}$ the mirror map. For
convenience, we also refer to the inverse $z_j(q)$ as the mirror map. Thus by
the construction above, the mirror map for an $s$-parameter family of Calab!
i-Yau mirror pairs can be regarde

\newsec{Construction of the Schwarzian equations}

We now discuss the construction of the differential
equation which governs the mirror map $z(t)$. We begin with an $n^{th}$ order
ODE of Fuchsian type:
\eqn\ode{Lf:=\left({d^n\over dz^n}+\sum_{i=0}^{n-1} q_i(z) {d^i\over
dz^i}\right)f=0}
($n$ will specialize to 2,3 and 4 later.) In particular,
the $q_i(z)$ are rational functions of $z$.
Let $f_1, f_2$ be two linearly independent solutions of this equation
and consider the ratio $t:=f_2(z)/f_1(z)$. Inverting this relation
(at least locally), we obtain $z$ as a function of $t$. Our goal is
to derive a polynomial ODE, in a canonical way, for $z(t)$.

We first perform a change of coordinates $z\rightarrow t$ on \ode~ and
obtain:
\eqn\odeb{\sum_{i=0}^n b_i(t) {d^i\over dt^i} f(z(t))=0}
where the $b_i(t)$ are rational expressions of the derivatives $z^{(k)}$
(including $z(t)$). For example we have $b_n(t)=a_n(z(t)) z'(t)^{-n}$.
It is convenient to put the equation in {\it reduced} form. We do a change of
variable $f=A g$, where $A=exp(-\int {b_{n-1}(t)\over n b_n(t)})$, and
multiply \odeb~ by ${1\over A b_n}$ so that it
becomes
\eqn\odec{\tilde{L}g:=\left({d^n\over dt^n} +\sum_{i=0}^{n-2} c_i(t) {d^i\over
dt^i}\right) g(z(t))=0}
where $c_i$ is now a rational expression
of $z(t),z'(t),..,z^{(n-i+1)}$ for
$i=0,..,n-2$. Now $g_1:=f_1/A$ and $g_2:=f_2/A=t g_1$
are both solutions to the equation \odec. In particular we have
\eqn\PQ{\eqalign{
P:=&\tilde{L}g_1=\left({d^n\over dt^n}+\sum_{i=0}^{n-2} c_i(t) {d^i\over
dt^i}\right) g_1=0\cr
Q:=&\tilde{L}(tg_1) -t\tilde{L}g_1=\left(n{d^{n-1}\over
dt^{n-1}}+\sum_{i=0}^{n-3} (i+1) c_{i+1}(t) {d^i\over dt^i}\right) g_1=0.}}
 Note that since $c_i$ is a rational expression of
$z(t),z'(t),..,z^{(n-i+1)}(t)$,
it follows that $P$ involves $z(t),..,z^{(n+1)}(t)$
while $Q$ involves only $z(t),..,z^{(n)}(t)$.
Eqns \PQ~ may be viewed as a coupled system of differential equations for
$g_1(t),z(t)$. Our goal
is to eliminate $g_1(t)$ so that we obtain an equation for just $z(t)$.
One way to construct this is as follows. By \PQ, we have
\eqn\system{\eqalign{
{d^i\over dt^i} P=&0,\ \ \ \ i=0,1,..,n-2,\cr
{d^j\over dt^j} Q=&0,\ \ \ \ j=0,1,..,n-1.}}
We now view \system~ as a homogeneous
{\it linear} system of equations:
\eqn\matrixform{\sum_{l=0}^{2n-2} M_{kl}(z(t),..,z^{(2n-1)}(t)){d^l\over
dt^l}g_1=0,\ \ \ k=0,..,2n-2,}
where $(M_{kl})$ is the following $(2n-1)\times(2n-1)$ matrix:
\eqn\M{\left(\matrix{
c_0&c_1&..&c_{n-2}&0&1&0&..&0\cr
c_0'&c_0+c_1'&..&c_{n-3}+c_{n-2}'&.&0&1&0&0\cr
 &\cdots&..&\cdots& & & &..&\cr
c_0^{(n-2)}&(n-2)c_0^{(n-3)}+c_1^{(n-2)}&
..&\cdots&.&.&.&0&1\cr
c_1&2c_2&..&(n-2)c_{n-2}&n&0&0&..&0\cr
c_1'&c_1+2c_2'&..&(n-3)c_{n-3}+(n-2)c_{n-2}'&0&n&0&..&0\cr
 &\cdots&..&\cdots& & &..&\cr
c_1^{(n-1)}&(n-1)c_1^{(n-2)}+2c_2^{(n-1)}&..&\cdots&.&
&.&0&n}\right)}
More precisely if we define the $1^{st}$ and $n^{th}$ ($n$ fixed)
row vectors to be
$(M_{1l})=(c_0,c_1,..,c_{n-2},0,1,0,..,0)$ and
$(M_{nl})=(c_1,2c_2,..,(n-2)c_{n-2},0,n,0,..,0)$ respectively, then the matrix
$(M_{kl})$ is given by the recursion relation:
\eqn\recursion{M_{k+1,l}=M_{k,l-1}+M_{k,l}',\ \ \ \
l=1,..,2n-1;\ \ k=1,..,n-2,n,..,2n-2.}
Thus the $(M_{kl})$ depends rationally on $z(t),..,z^{(2n-1)}(t)$. Since $g_1$
is nonzero, it follows that
\eqn\det{det\left( M_{kl}(z(t),..,z^{(2n-1)}(t))\right)=0.}
We call this the $n^{th}$ Schwarzian equation associated with \ode. Note that
by suitably clearing denominators, the equation becomes a $(2n-1)^{st}$ order
polynomial ODE for $z(t)$ with constant coefficients.
It is clear that this equation depends only on the data $q_i(z)$ we began
with.

\newsec{The analytic solutions to the Schwarzian equation}

It was already known to H.A. Schwarz that all the solutions to his equation
(n=2) can be constructed from the original 2nd order ODE. In this section, we
wish to generalizes his result to the other higher Schwarzian equations. We
will focus on those which arise only from cases in which the data $q_i(z)$ come
from the Picard-Fuchs equation of a smooth Calabi-Yau hypersurface (or complete
intersections) in a weighted projective space. We will show that there is
exactly 1-parameter family of single-valued analytic solutions, on a disk, to
the Schwarzian equation.
It is instructive to first go back to the n=2 case (elliptic curves). While
Schwarz's treatment in this case focus on the solution ratios (triangle
functions), it is less direct for studying their inverses. Instead we will
study the inverses directly. This will have the advantages of a) exhausting all
the analytic solutions directly; b) seeing manifestly that the solutions have
polynomial dependence on the parameters; and c) generalizing immediately to
higher Schwarzian equations.

\subsec{n=2 Schwarzian equation}
\subseclab\twose

We begin with a Fuchsian equation of the general form:
\eqn\pfA{\left(\Theta^2 - \delta z (\Theta+a) (\Theta +b)\right)f(z)=0}
where the $\delta,a,b$ are constants with $\delta\neq0$ (later we will restrict
to rational numbers with $0<a,b<1$), and $\Theta=z{d\over dz}$. The
Picard-Fuchs equations for elliptic curves in a weighted projective space (For
the general form of the Picard-Fuchs equation see  \hktyI\hktyII.) is of this
form where $z$ is a suitable coordinate for the complex structure moduli space
of the curves. The periods of the curves are given by the solutions to \pfA.
This equation is projectively equivalent to a hypergeometric equation:
\eqn\hge{z(1-z) y''+(1-(a+b+1)z)y'-aby=0}
ie. there exists a function $A$ and a Mobius transformation
$z\rightarrow{\alpha z+\beta\over \gamma z+\delta}$ such that $f=Ay$ transforms
\pfA~ into \hge. (Take $z\rightarrow z/\delta$.) The Schwarzian equation
associated to \pfA~ is
\eqn\seA{2Q(z(t)) {z'}^2 +\{z,t\}=0}
where $\{z,t\}={z'''\over z'}-{3\over 2}({z''\over z'})^2$ and
\eqn\potential{Q(z)={1+2 (-1 + a + b - 2 a b) \delta z + (-1 + a - b) (-1 - a +
b) \delta^2 z^2\over 4z^2(1-\delta z)^2.}}
The Schwarzian equation associated with \hge~ is \seA~ with $\delta=1$. For
convenience, we let $\delta=1$ for now. To restore the generality, one simply
replace $z$ by $\delta z$ below.
It was known to H.A. Schwarz that the every solution $z(t)$ to \seA is the
inverse of a ratio (known as a Schwarz triangle function of type
$s(0,b-a,1-a-b;z)$) of two hypergeometric functions which solve \hge. In this
note, we will only be interested in {\it solutions $z(q)$ to \seA which are
analytic in some disk $|q|<R$ where $q=e^{2\pi i t}$}. It turns out that every
analytic solution $z$ must have either $z(q=0)=0$ or $1$. We will only consider
those with $z(0)=0$. They can be obtained as follows. Eqn \pfA has a unique
power series solution $f_1(z)$ with $f_1(0)=1$, namely
$f_1(z)=\sum_{n\geq0}c_n z^n$ where $c_n={\Gamma(n+a)\Gamma(n+b) \over
\Gamma(a)\Gamma(b)\Gamma(n+1)^2}$. There is also a unique solution of the form
$f_2(z)=log(z) f_1(z) +\sum_{n\geq1} d_n z^n$. Note that both
the $c_n,d_n$ depends polynomially on the parameters $a,b$.
Let $t={f_2(z)\over 2\pi i f_1(z)}$. Then
\eqn\ratio{q=z\ exp(\sum d_n z^n/ \sum c_n z^n).}
This defines an invertible analytic map from a disk $|z|<S$ to some $|q|<R$
sending $0$ to $0$. The inverse map $z_f(q)$ therefore gives a particular
analytic solution to \seA, which we will call {\it the fundamental solution}.
Note also the the Fourier coefficients of $z_f(q)$ also depend polynomially on
$a,b$. (Thus the above construction makes sense
with no restriction at all on the values of $a,b$.) It is convenient to
transform \seA by $q=e^{2\pi it}$ so that the equation now has $q$ as an
independent variable and $z(q)$ dependent variable.

We will argue that $\{z_f(kq)\}_{k\in {\bf C}^\times}$ exhausts all analytic
solutions $z(q)$ with $z(0)=0$ and ${dz(0)\over dq}\neq0$, ie. every such
solution can be obtained from the fundamental one by scaling $q$. First note
that scaling $q$ by $k$ corresponds to a translation $t\rightarrow t+\beta$ for
some $\beta$. But we know that \seA is invariant under Mobius transformations
on $t$.
This shows that each $z_f(kq)$ is an analytic solution to \seA. It remains to
show that if $z(q)=\sum_{n\geq1} a_n q^n$ is an analytic solution to \seA with
$a_1\neq0$, then the $a_n$ are determined by $a_1$. Since \seA is 3rd order
and is linear in $d^3 z\over dq^3$, $a_n$ for $n=3,4,...$ are determined by
$a_1$ and $a_2$. But a simple computation using \seA gives
\eqn\dumb{a_2=(2ab-a-b) a_1.}

As argued, the Fourier coefficients of $z_f(kq)$ are polynomials in $a,b,k$.
The first few terms of $z_f(q)$ are
\eqn\dumb{\eqalign{
z_f(q)&=q + (-a - b + 2 a b) q^2 \cr & -
 (a - 5 a^2  + b - 11 a b + 19 a^2  b - 5 b^2  + 19 a b^2  - 19 a^2  b^2 )q^3/4
\cr &-(6 a - 45 a^2  + 93 a^3  + 6 b - 94 a b + 408 a^2  b - 515 a^3  b - 45
b^2  \cr &+ 408 a b^2  - 1116 a^2  b^2  + 987 a^3  b^2  + 93 b^3  - 515 a b^3
+
987 a^2  b^3  - 658 a^3  b^3 )q^4/54+ ...
}}
This is obtained by either inverting \ratio  or by directly solving \seA.

\subsec{n=3 Schwarzian equation}
\subseclab\threese

It is now clear how to generalize the above. Consider the following 3rd order
ODE of Fuchsian type:
\eqn\pftemp{\left(\Theta^3 - z(\sum_{i=0}^3 r_i\Theta^i)\right)f(z)=0.}
where the $r_i$ are constants with $r_3\neq0$.
The Picard-Fuchs equations for (a 1-parameter deformations of) smooth K3
hypersurfaces or complete intersections in weighted projective space is of this
form (see \hktyI\hktyII) where the $r_i$ are integer valued and satisfy some
restriction given as follows. There is a ``uniformizing'' coordinate $\tau$ in
which \pftemp~ becomes ${d^3 \tilde{f} \over d\tau^3}=0$. This implies that the
following three possibilities:
\eqn\dumb{\eqalign{
a)&\ r_0=r_1=r_2=0;\cr
b)&\ 3r_0=r_1=r_2=3r_3;\cr
c)&\ r_0={2 r_1 - r_3\over 4},\ \ r_2={3 r_3\over 2}.
}}
Cases a) and b) turns out to be projectively equivalent, ie. they both result
in the same reduced form for \pftemp:
\eqn\dumb{\left( {d^3\over dz^3} + {1\over 4z^2}{d\over dz} -
{1\over 4z^3}\right)\tilde{f}(z)=0.}
This case will not arise in our discussion below. Thus from now on we impose c)
and hence \pftemp~ becomes, after simplication:
\eqn\pfB{\left(\Theta^3 - r_3 z (\Theta+{1\over 2})(\Theta^2 +\Theta +
{r_1\over r_3}-{1\over 2})\right)f(z)=0.}
According to our general construction. The Schwarzian equation associated with
\pfB~ is then (see eqn \det)
\eqn\schwarzIII{
det(M_{kl}):=27 c_0^2 + 4c_1^3-18c_1 c_0'-3{c_1'}^2+ 6c_1 c_1''=0.}
where
\eqn\ccexpression{\eqalign{
c_1(t) :=& q_1(z)\,{{z'}^2} + 2\{z,t\} \cr
c_0(t) :=& q_0(z)\,{{z'}^3} + q_1(z)\,z'\,z'' +
   {{3\,{{z''}^3}}\over {{{z'}^3}}} \cr &-
   {{4\,z''\,z^{(3)}}\over {{{z'}^2}}} + {{z^{(4)}}\over {z'}}
\cr
q_1(z):=&{4 - 4 r_1z - 2 r_3 z + 4 r_1r_3 z^2 + r_3^2 z^2\over 4 z^2  (-1 +
r_3z)^2}\cr
q_0(z):=&{1\over 2} {dq_1(z)\over dz}.
}}
It is easy to check that $c_0(t)={1\over 2}c_1'(t)$. This simplifies
\schwarzIII to
\eqn\seB{15 {c_1'}^2 + 16 c_1^3 - 12 c_1 c_1'' =0.}
Note that this simplification is a direct consequence of imposing c) above.
This n=3 Schwarzian equation will be useful for understanding the mirror map
for K3 surfaces (see below).

We now construct all analytic solutions $z(q)$ to \seB~ with $z(0)=0$ and
${dz(0)\over dq}\neq0$. The situation here is completely analogous to the n=2
case. Eqn \pfB~ has a unique power series solution $f_1(z)=\sum_{n\geq0}c_n
z^n$ with $f_1(0)=1$. There is also a unique solution of the form
$f_2(z)=log(z) f_1(z) +\sum_{n\geq1} d_n z^n$. The $c_n,d_n$ are polynomials of
the parameter $r_1,r_3$. If we let $t={f_2(z)\over 2\pi i f_1(z)}$, then
\eqn\dumb{q:=e^{2\pi i t}=z\ exp(\sum d_n z^n/ \sum c_n z^n)}
defines an invertible analytic map from a disk $|z|<S$ to some $|q|<R$. The
inverse map $z_f(q)$ therefore gives a particular analytic solution to \seB. By
an argument similar to the n=2 case, we conclude that
 $\{z_f(kq)\}_{k\in {\bf C}^\times}$ exhausts all such analytic solutions. The
Fourier coefficients of the fundamental solution $z_f(q)$ can be computed using
\seB:
\eqn\dumb{\eqalign{
z_f(q)=&q + (2 r_1-3r_3)q^2/4 + (76 r_1^2  - 212 r_1 r_3 + 135 r_3^2 )
q^3/256\cr
&+ (2632 r_1^3  - 10468 r_1^2  r_3 + 12862 r_1 r_3^2  - 5007 r_3^3 ) q^4/13824
\cr
&+(1806544 r_1^4  - 9219424 r_1^3  r_3 \cr &+ 16526488 r_1^2  r_3^2  -
 12589560 r_1 r_3^3  + 3479157 r_3^4 ) q^5  / 14155776 + ...
}}

The above result has an interesting consequence: {\it every analytic solution
to \seB is a solution to:}
\eqn\seBsolve{c_1:=q_1(z)\,{{z'}^2} + 2\{z,t\}=0.}
To prove this, note that \seBsolve~ is an n=2 Schwarzian equation \seA~ with
$Q(z)={1\over 4}q_1(z)$. Associated to it are (in general) four
2nd order Fuchsian equation of the type \pfA~ with parameters $\delta=r_3$ and
$a,b$ satisfying
\eqn\dummy{\eqalign{
2ab-a-b&={r_1\over 2 r_3}-{3\over 4}\cr
(a-b)^2&=-{r_1\over r_3}+{3\over 4}.
}}
(Note that if $(a,b)$ solve \dummy, so do $(b,a)$ and $(1-a,1-b)$.) By our
previous result on n=2 Schwarzian equation, \seBsolve~ has a family of
solutions $\{z_f(kq)\}_{k\in {\bf C}^\times}$. Obviously they are solutions to
\seB~ as well. Thus
by our result on n=3 Schwarzian equation, this family exhaust the solutions to
\seB. Thus we have {\it effectively reduced the n=3 Schwarzian equation to the
n=2 equation.}

\subsec{n=4 Schwarzian equation}
\subseclab\fourse

For completeness, we briefly discuss the n=4 case even though we will not be
applying the result later. The situation here is quite similar to the n=3 case,
except the last part above. Consider the following 4rd order ODE:
\eqn\pftemp{\left(\Theta^4 - z(r_4\Theta^4 + 2 r_4 \Theta^3 + r_2 \Theta^2
+(r_2-r_4)\Theta +r_0)\right)f(z)=0.}
where the $r_i$ are constants with $r_4\neq0$.
The Picard-Fuchs equations for certain 1-parameter family of Calabi-Yau
threefolds in weighted projective space is of this form where the $r_i$ are
integers. According to our general construction, the Schwarzian equation
associated with \pfB~ is then (see eqn \det)
\eqn\schwarzIV{\eqalign{
det(M_{kl}):=&16\,{{c_2 }^4}\,c_0  - 128\,{{c_2 }^2}\,{{c_0 }^2} +
   256\,{{c_0 }^3} + 4\,{{c_2 }^3}\,{{c_2 '}^2} \cr+
&   240\,c_2 \,c_0 \,{{c_2 '}^2} - 15\,{{c_2 '}^4} -
   144\,{{c_2 }^2}\,c_2 '\,c_0 ' -
   448\,c_0 \,c_2 '\,c_0 ' \cr+
&   256\,c_2 \,{{c_0 '}^2} - 8\,{{c_2 }^4}\,c_2 '' +
   128\,{{c_0 }^2}\,c_2 '' -
   48\,c_2 \,{{c_2 '}^2}\,c_2 '' \cr+
&   48\,c_2 '\,c_0 '\,c_2 '' +
   12\,{{c_2 }^2}\,{{c_2 ''}^2} -
   48\,c_0 \,{{c_2 ''}^2} + 32\,{{c_2 }^3}\,c_0 '' \cr-
&   128\,c_2 \,c_0 \,c_0 '' +
   48\,{{c_2 '}^2}\,c_0 '' +
   32\,{{c_2 }^2}\,c_2 '\,c_2 ^{(3)} +
   64\,c_0 \,c_2 '\,c_2 ^{(3)} \cr-
&   96\,c_2 \,c_0 '\,c_2 ^{(3)} +
   8\,c_2 \,{{c_2 ^{(3)}}^2} -
   8\,{{c_2 }^3}\,c_2 ^{(4)} +
   32\,c_2 \,c_0 \,c_2 ^{(4)} -
   12\,{{c_2 '}^2}\,c_2 ^{(4)}=0\cr
}}
where
\eqn\ccrr{\eqalign{
c_2(t):=& q_2(z){z'}^2 + 5\{z(t),t\}_2 \cr
c_0(t):=& q_0(z){z'}^4 +
   {3\over 2} {dq_2(z)\over dz}{z'}^2 z'' -
   {3\over 4} q_2(z) {z''}^2 -
   {{135{z''}^4}\over {16{z'}^4}} \cr+
  & {3\over 2} q_2(z) z' z^{(3)}+
   {{75{z''}^2 z^{(3)}}\over {4{z'}^3}} -
   {{15{z^{(3)}}^2}\over {4{z'}^2}} -
   {{15z'' z^{(4)}}\over {2{z'}^2}} +
   {{3z^{(5)}}\over {2z'}} \cr
q_2(z):=&{5 - 2 z r_2 - 4 z r_4 + 2 z^2  r_2 r_4 + 2 z^2  r_4^2\over
2 z^2  (-1 + z r_4)^2}\cr
q_0(z):=&(81 - 16 z r_0 - 12 z r_2 - 280 z r_4 + 48 z^2  r_0 r_4 +
44 z^2  r_2 r_4 + 340 z^2  r_4^2  - 48 z^3  r_0 r_4^2 \cr & - 64 z^3  r_2 r_4^2
- 128 z^3  r_4^3  + 16 z^4  r_0 r_4^3  + 32 z^4  r_2 r_4^3  + 32 z^4  r_4^4 ) /
(16 z^4  (-1 + z r_4)^4 )
}}

Once again, the set of analytic solutions to \schwarzIV with
$z(q=0)=0$ and ${dz(0)\over dq}\neq0$ is of the
form $\{z_f(kq)\}_{k\in{\bf C}^\times}$, where $z_f(q)$ is the fundamental
solution whose Fourier coefficients are polynomials in
the data $(r_4,r_2,r_0)$. The construction is almost identical to
that for n=2,3. The first few coefficients $z_f(q)$ are given by:
\vfill
\eject
\eqn\seriesIV{\eqalign{
z_f(q) &= q + {q^2}\, (4\,r_0 - r_2 + r_4) \cr &+
   {q^3}\,\left( 163\,{{r_0}^2} - 84\,r_0\,r_2 + 11\,{{r_2}^2} +
         90\,r_0\,r_4 - 24\,r_2\,r_4 + 13\,{{r_4}^2} \right)/8 \cr&+
   {q^4}\,\left( 112798\,{{r_0}^3} - 88840\,{{r_0}^2}\,r_2 +
         23610\,r_0\,{{r_2}^2} - 2115\,{{r_2}^3} \right.\cr&+
         98969\,{{r_0}^2}\,r_4 - 53280\,r_0\,r_2\,r_4 +
         7245\,{{r_2}^2}\,r_4 + 29958\,r_0\,{{r_4}^2} \cr&-\left.
         8253\,r_2\,{{r_4}^2} + 3123\,{{r_4}^3} \right)/{972} \cr &+
   {q^5}\,\left( 702344153\,{{r_0}^4} - 747165436\,{{r_0}^3}\,r_2 +
         300965204\,{{r_0}^2}\,{{r_2}^2} -54365280\,r_0\,{{r_2}^3}\right.\cr &+
         3713328\,{{r_2}^4} + 853805796\,{{r_0}^3}\,r_4 -
         694587576\,{{r_0}^2}\,r_2\,r_4 +
         189938400\,r_0\,{{r_2}^2}\,r_4 \cr&- 17448192\,{{r_2}^3}\,r_4 +
         400090708\,{{r_0}^2}\,{{r_4}^2} -
         220959840\,r_0\,r_2\,{{r_4}^2} +
         30726432\,{{r_2}^2}\,{{r_4}^2} \cr&+ \left. 85542240\,r_0\,{{r_4}^3} -
        24023808\,r_2\,{{r_4}^3} + 7032240\,{{r_4}^4} \right)/
     {995328} \cr & + {q^6}\,\left( 2322744173252\,{{r_0}^5} -
         3118439852795\,{{r_0}^4}\,r_2 +
         1688073976516\,{{r_0}^3}\,{{r_2}^2} \right. \cr&-
         460304004892\,{{r_0}^2}\,{{r_2}^3} +
         63195785760\,r_0\,{{r_2}^4} - 3493177200\,{{r_2}^5} \cr&+
         3628517760463\,{{r_0}^4}\,r_4 -
         3959225703216\,{{r_0}^3}\,r_2\,r_4 +
         1631456931956\,{{r_0}^2}\,{{r_2}^2}\,r_4 \cr&-
         300755456640\,r_0\,{{r_2}^3}\,r_4 +
         20919798000\,{{r_2}^4}\,r_4 +
         2319777533708\,{{r_0}^3}\,{{r_4}^2} \cr &-
         1926736953572\,{{r_0}^2}\,r_2\,{{r_4}^2} +
         536717533440\,r_0\,{{r_2}^2}\,{{r_4}^2} -
         50124477600\,{{r_2}^3}\,{{r_4}^2} \cr &+
         757958047180\,{{r_0}^2}\,{{r_4}^3} -
         425557676160\,r_0\,r_2\,{{r_4}^3} +
         60050628000\,{{r_2}^2}\,{{r_4}^3} \cr &+ \left.
         126449580000\,r_0\,{{r_4}^4} - 35961865200\,r_2\,{{r_4}^4} +
         8609094000\,{{r_4}^5} \right)/{518400000} + ...
}}
It is interesting to note that this series appears to be integral
when evaluated at certain integral points. For example, if the data
$(r_4,r_2,r_0)=(3125,4375,120)$ corresponding to the quintic threefold
in ${\bf P}^4$, then \seriesIV~ becomes:
\eqn\dumb{z(q)= q - 770 q^2 + 171525 q^3 - 81623000 q^4 - 35423171250 q^5 -
   54572818340154 q^6-...}
This integrality phenomenon apparently continues to hold in other
examples as has been previously observed.

At this point, one might wonder if $z_f$ would satisfy an n=2
Schwarzian equation \seA~ as in the previous case, ie. is there
a rational function $Q(z)$ similar to \potential such that \seriesIV~
solves the equation \seA? The answer is no in general. In fact,
we have checked that in many cases of a 1-parameter family of Calabi-Yau
threefolds, such a $Q(z)$ doesn't exist.(cf. \cdgp section 2.) However the
mirror map $z_f$ of such a threefold
does satisfy an equation similar to \seA, but with some
``quantum'' corrections, namely:
\eqn\sequantum{
2 Q(z) {z'}^2 +\{z,t\} = {2\over 5} y''-{1\over 10} {y'}^2
}
where $Q(z)={1\over10} q_2(z)$ (eqn \ccrr) and $y=log K(t)$, $K$ being the
quantum Yukawa coupling (also written as $\partial^3F$). Thus the right hand
side of \sequantum should be thought of as a ``quantum'' correction to the
classical equation \seA.

\subsec{Remarks}
\subseclab\remarksI

1. In the last section, we have seen that the Schwarzian equations can
essentially
be solved in terms of certain hypergeometric functions. Moreover the analytic
solution is essentially unique. In fact, it follows easily from the previous
discussion that in each case n=2,3 or 4, the fundamental solution to the
$n^{th}$ Schwarzian equation is the unique analytic solution (single-valued)
$z_f(q)$ with $z_f(0)=0, {dz_f(0)\over dq}=1$. This is exactly the so-called
mirror map. Thus {\it the mirror map can be characterized by means of an ODE
and analyticity.} We also find that there is a similar characterization for the
quantum coupling $K(t)$ in a number of examples.

2. Even though we have obtained only the singled-valued analytic solutions to
the Schwarzian equations in the disk $|q|<R$, the results actually allows us to
classify certain multi-valued solutions as well. For simplicity, let's focus on
the n=2 case. What we say here will hold true for n=3,4 as well. For example,
given any positive constant $\alpha$, one can easily classify the analytic
solutions $z(q)$ to \seA such that $z(q)\sim kq^\alpha$ as $q\rightarrow0$ for
some constant $k$. First note that there is a 1-parameter family of such
solutions, namely $z_f(kq^\alpha)$ where $z_f(q)$ is our fundamental solution.
This follows immediately from Mobius invariance of \seA. This family also
exhausts all such solutions.

3. We studied the Schwarzian equations associated with Fuchsian equations of
special type which arises as the Picard-Fuchs equations for certain Calabi-Yau
manifolds. It is clear that one can study this Schwarzian equations without
reference to special geometry, in which case, some of our assumption can be
weaken. For example, we have checked that the uniqueness of analytic solutions
to these equations continues to hold true under much weaker assumptions on the
form of the associated Fuchsian equations. (For instance, take any Fuchsian
equation whose indicial equation is maximally degenerate, ie. whenever the
differential operator takes the form $\Theta^n-p(z,\Theta)$ where $p$ is a
polynomial of degree at most $n$ in $\Theta$, with $p(0,\Theta)=0$.)

4. We have shown that the mirror map in the case of K3 surfaces satisfies an
n=2 Schwarzian equation, as in the case of elliptic curves. This is the first
hint that there might be some relationship between K3 surfaces and elliptic
curves.
It also says that the mirror map is nothing but the inverse of a Schwarz
triangle function. An even more interesting hint comes from the following
numerical experiment. If we consider a 1-parameter family of sextic
hypersurfaces in the weighted projective space ${\bf P}^3[1,1,1,3]$ and compute
the Fourier series of the mirror map, we get
\eqn\dumbseries{z(q)=q - 744 q^2  + 356652 q^3  - 140361152 q^4  + 49336682190
q^5  -16114625669088 q^6+...}
The observant reader would have realized that the coefficient 744 is the
constant term in the J-function times 1728. In fact if we compute $1/z(q)$, we
get exactly the first 6 coefficients of the 1728J. We will prove that
$1728J=1/z(q)$. This will be another confirmation that the mirror map for a
Calabi-Yau variety is integral.

\newsec{Transcendence of the mirror map over ${\bf Q}(J)$}

In this section analytic functions are assumed to be defined on some disk
$|q|<R$. Let ${\bf Q}(f)$ be the field generated by the analytic functions
$f(q)$ over the rationals ${\bf Q}$. Two analytic functions $f(q),g(q)$ are
called commensurable if there exists a nontrivial polynomial relation (with
coefficients in ${\bf Q}$) $P(f(q),g(kq^\alpha))=0$ for some rational numbers
$k,\alpha$. The statement that $f(q)$ and $g(q)$ are commensurable is
equivalent to that $g(kq^\alpha)$ is algebraic over the field ${\bf Q}(f)$.
Note also that commensurability is an equivalence relation.

We will be mainly concerned with the question of commensurability of an
analytic function $f(q)$ with the J-function. Perhaps the simplest example of a
function commensurable to $J$ is the so-called elliptic modular function (see
below) $\lambda$. It bears the following well-known modular relation to $J$:
\eqn\dumb{J={4\over27}{(1-\lambda+\lambda^2)^3\over\lambda^2(1-\lambda)^2}.}
The function $\lambda$ will show up again in the next section.

In this section, we will prove that the mirror map $z(q)$ for K3 surfaces
modelled in various weighted projective spaces is commensurable with $J$.
Before discussing K3 surfaces, it is useful to recall some known examples of
mirror maps which are commensurable with $J$
\ref\VerlindeWarner{E. Verlinde and N. Warner, \plb269(1991)96}
\ref\KlemmTheisenSchmidt{A. Klemm, S. Theisen and M. Schmidt ,
\ijmpa7(1992)6215.} (see also \klryA~ section 3.1).  For both the elliptic
curves and the K3 surfaces, we will use the commensurability relation to prove
that the mirror maps in those cases are integral.

\subsec{Elliptic curves}
\subseclab\elliptic

In the following table,
we have four realizations of a 1-parameter family of elliptic curves
as hypersurfaces (or complete intersections) in weighted projective spaces
${\bf P}^2(1,1,1),{\bf P}^2(1,1,2)$,${\bf P}^2(1,2,3)$ and ${\bf
P}^3(1,1,1,1)$.
$$
\vbox{\offinterlineskip\tabskip=0pt
\halign{\strut\vrule#
&\hfil~$#$
&\vrule#
&~~$#$~~\hfil
&~~$#$~~\hfil
&~~$#$~~\hfil
&\vrule#\cr
\noalign{\hrule}
& && {\rm deformations} &{\rm  diff.\,\, operator}& 1728 J(z)& \cr
\noalign{\hrule}
&1. && x_1^3+x_2^3+x_3^3+z^{-1/3}x_1x_2x_3=0&
\theta^2- 3 z ( 3 \theta+2)(3\theta+1)&
\displaystyle{(1+ 216z)^3\over z (1-27 z)^3} &\cr
&2. && x_1^4+x_2^4+x_3^2+z^{-1/4}x_1x_2x_3=0&
\theta^2- 4 z ( 4 \theta+3)(4\theta+1)&
\displaystyle{(1+ 192z)^3\over z (1-64 z)^2} &\cr
&3. && x_1^6+x_2^3+x_3^2+z^{-1/6}x_1x_2x_3=0&
\theta^2- 12 z ( 6 \theta+5)(6\theta+1)&
\displaystyle{1\over z (1-432 z)} &\cr
&4. && x_1^2+x_2^2+z^{-1/4}x_3x_4=0&
\theta^2- 4z ( 2 \theta+1)^2&
\displaystyle{(1 + 224 z + 256 z^2)^3\over z (-1 + 16 z)^4} &\cr
& && x_3^2+x_4^2+z^{-1/4}x_1x_2=0&
 & &\cr
\noalign{\hrule}}
\hrule}$$
Thus in those four examples, the mirror maps $z$ are algebraic over the field
${\bf Q}(J)$. The relations between $z$ and $J$ in the some of these examples
are derived in \klryA~ by using the Weierstrass model for the elliptic curves.
We will instead illustrate the proof using a slightly different approach which
will prove useful in the case of K3 surfaces. Namely, we will effectively use
the results on the uniqueness of analytic solutions to the associated n=2
Schwarzian equations. We will consider example 4 because it has some
interesting connection with elliptic functions and it hasn't been treated in
\klryA. Our goal is to prove the relation between $z$ and $J$ in example 4
above. (We also have similar proofs for all the other cases above.)

Recall that $J(t):={g_2^3(t)\over\Delta(t)}$ where $\Delta(t)$ is the
discriminant of the Weierstrass elliptic curve $W$ with period ratio $t$:
\eqn\dumb{W:\ y^2=4x^3-g_2(t)x-g_3(t).}
The periods $\Omega$ of $W$ are solutions to the Picard-Fuchs equation:
\eqn\universal{{d^2\Omega \over dJ^2} +{1\over J}{d\Omega\over dJ} +{31 J-
4\over 144 J^2(1-J)^2 }\Omega=0.}
Thus $J$ is a solution to the associated Schwarzian equation \seA~ with:
\eqn\dumb{Q(z):={32 - 41 z+36z^2\over 144 z^2(-1 + z)^2  }.}
The function $J(q)$ ($q:=e^{2\pi it}$) can be characterized as the unique
meromorphic solution to the Schwarzian equation on a disk $|q|<R$ with the
leading behavior $J(q)\sim {1\over 1728 q}$ near $0$. It is more convenient to
work with $z_W(q):={1\over 1728J(q)}$ which satisfies \seA~ with:
\eqn\potentialW{Q(z)=Q_W(z):=
{1 - 1968 z + 2654208 z^2\over 4 z^2  (-1 + 1728 z)^2}}
which is in the standard form \potential. Since $z_W$ has the leading behavior
$z_W(q)\sim q$, it follows that (section \twose) $z_W$ is the fundamental
solution to \seA~ with \potentialW.

Now consider the mirror map $z_X$ for example 4 above. This is the fundamental
solution to the Schwarzian equation \seA with:
\eqn\potentialX{Q(z)=Q_X(z):={1 - 16 z + 256 z^2\over 4 z^2  (-1 + 16 z)^2.}}
We want to prove that
\eqn\zetaq{z_W={z_X (-1 + 16 z_X)^4\over(1 + 224 z_X + 256 z_X^2)^3}.}
Note that the right hand side, which we will call $\zeta$, has the correct
leading behavior. By subsituting $\zeta$ into \seA with \potentialW and use the
fact the $z_X$ satisfies \seA~ with \potentialX, we find that indeed \seA~
holds. By uniqueness of analytic solution to the Schwarzian equation we
conclude that $\zeta=z_W$. We remark that one can also prove \zetaq~ simply by
applying the variable change $z\rightarrow {z (-1 + 16 z)^4\over(1 + 224 z +
256 z^2)^3}$ to the (reduced) Fuchsian equation $f''(z)+Q_W(z)f(z)=0$ and see
that you get
the new reduced Fuchsian equation $\tilde{f}''(z)+Q_X(z)\tilde{f}(z)=0$. It
follows that their respective associated Schwarzian equations must transform
from one to another under the same variable change.

\subsec{Integrality of $z_X$ and the elliptic modular function $\lambda$}

There is an amusing connection which we should point out between $z_X$ and
$\lambda$. The latter is defined to be the inverse function of the Schwarz
triangle function \ref\bateman{H. Bateman, Higher Transcendental Functions,
McGraw Hill, 1953}:
\eqn\dumb{t=S(0,0,0;z):={\sqrt{-1}F({1\over2},{1\over2},1;1-z)\over
F({1\over2},{1\over2},1;z)}}
where the $F$ are hypergeometric functions with
$(a,b,c)=({1\over2},{1\over2},1)$. As a result we can also characterize
$\lambda$ as the unique analytic solution, with the leading behavior
$\lambda(q)\sim 16q^{1\over2}$, to the
Schwarzian equation \seA (associated to \hge) with:
\eqn\dumb{Q(z)=Q_E(z):={1 - z + z^2\over 4 z^2  (-1 + z)^2}.}
The function $\lambda$ has a q-series expansion:
\eqn\dumb{\lambda(q)=16q^{1/2}-128q+704q^{3/2}-3072q^2+11488q^{5/2}
-38400q^3 + 117632q^{7/2} -...}
{}From its expression in terms of theta functions, it is known that $\lambda$
has integral Fourier coefficients divisible by 16.
It is also well-known that $\lambda$ defines a modular function of level 2 on
the upper half-plane and it is related to $J$ by:
\eqn\lambdaq{J={4\over27}{(1-\lambda+\lambda^2)^3\over\lambda^2(1-\lambda)^2}.}
(This relation can also be proved using the approach we used to prove \zetaq.)
 Since commensurability is an equivalence relation, it follows from \zetaq~ and
\lambdaq~ that $z_X(q)$, $\lambda(q^2)$ are commensurable also. In fact, using
the method above, one can easily prove that
\eqn\lambdazX{\lambda(q^2)=16 z_X(q).}
Since the Fourier coefficients on the left hand side are integers divisible by
16, it follows that $z_X$ also has integral Fourier coefficients. We should
also point out that combining the relations \zetaq, \lambdaq~ and \lambdazX~,
we get a rather perculiar identity for $\lambda$:
\eqn\perculiar{
{16\lambda(q^2) (-1 + \lambda(q^2))^4\over(1 + 14 \lambda(q^2) +
\lambda(q^2)^2)^3}=
{\lambda(q)^2(1-\lambda(q))^2\over(1-\lambda(q)+\lambda(q)^2)^3}.}
It also says that you can write $J$ in terms of $\lambda$ in two different
ways.

\subsec{Remarks}
\subseclab\remarksII

1. The above approach for studying the commensurability of mirror maps also
applies easily to examples 1-3 in the previous section.

2. In the last example we studied, it is less clear how to deduce, from the
relation \zetaq, the integrality of $z_X$ from the known integrality property
of $z_W$. But upon relating $z_X$ to $\lambda$ via \lambdazX, this property
becomes immediately clear. The upshot of this is that using the arithmetic
properties of any particular one mirror map, such as $z_W$, alone may only give
partial information about mirror maps commensurable to it. One should instead
use other series, such as $\lambda$, in the same commensurability class to help
obtain further information about other members in the same class. The lesson is
that the larger the commensurability class, the more arithmetical information
we can get about its members because every pair of members are related by some
modular relations.

3. The key steps we used repeatedly above to prove commensurability of two
fundamental solutions are:

a) identify the appropriate modular relation;

b) relate their corresponding Schwarzian equations (or equivalently their
reduced Fuchsian equations) by a change of variables using the modular
relation;

c) use analyticity (asymptotic behavior) and uniqueness of solutions to the
Schwarzian equation;

This idea will continue to work well, as we will see, in the case of
fundamental solutions arising from the Picard-Fuchs equations for K3 surfaces.

4. Finally, we note that in e.g. 3 of elliptic curves above, we have the
relation, for $Im\ t>>0$
\eqn\wj{w(t)={j(t)+(j(t)(j(t)-1728))^{1/2}\over 2} }
where $w=1/z$,$j=1728J$. We claim that $w(t)$ admits a double-valued analytic
continuation in the upper half plane $Im\ t>0$. This implies in particular that
the Fourier series of the mirror map $z=1/w$ has radius of convergence strictly
less than 1 -- a fact that is not obvious from the construction of $z$. If we
denote by $z(t),\tilde{z}(t)$ the two branches of $z$, we see that
$z(t)+\tilde{z}(t)={1\over432}$. Thus the two branches differ by an affine
transformation $x\rightarrow -x+{1\over432}$.

To prove our claim, let's recall some properties of the modular function
$j(t)$.
We know that $j(t)$ is a single-valued function in $Im\ t>0$. Thus \wj~ implies
that $w(t)$ admits an analytic continuation which is {\it at most}
double-valued. Also for $\rho:={1\over 2}+{i\sqrt{3}\over 2}$ we have
$j(\rho)=0$. In a small punctured disk centered at $\rho$, $(j(t)-1728)^{1/2}$
is single-valued. If we move around a small loop enclosing $\rho$ in that disk,
$j(t)$ will move around a small loop (3 times) enclosing $0$ in the j-plane. It
follows that $(j(t)(j(t)-1728))^{1/2}$ is necessarily double-valued in that
disk, implying that $w(t)$ is {\it at least} double-valued.

\subsec{K3 surfaces: $X_z:~~ x_1^6+x_2^6+x_3^6+x_4^2 + z^{-1/6} x_1 x_2
x_3x_4=0$}

We now apply what we learned in section \threese~ to this family of K3
surfaces.
The Picard-Fuchs equation for the above 1-parameter family of K3 hypersurfaces
in ${\bf P}^3[1,1,1,3]$ is given by:
\eqn\dumb{\left(\Theta^3  - 8z (6 \Theta+1) (6 \Theta+3)
(6 \Theta+5)\right)f(z) =0}
which is of the form \pfB~ with $(r_3,r_1)=(1728,1104)$. By the results in
section \threese, the analytic solutions of the associated n=3 Schwarzian
equation \seB can be obtained from the n=2 Schwarzian equation \seA~ with:
\eqn\sexticQ{Q(z)=Q_X(z):=
{1 - 1968 z + 2654208 z^2\over 4 z^2  (-1 + 1728 z)^2}.}
In particular the mirror map $z_X$ for our K3 surfaces $X$ which is the
fundamental solution to \seB, is now the fundamental solution to \seA~ with
\sexticQ. But observe that $Q_X(z)$ is identical to $Q_W(z)$ given in
\potentialW. It follows that our mirror map is given by:
\eqn\dumb{z_X=z_W:={1\over 1728J}.}
This also proves, in particular, that $z_X$ also has an integral Fourier
coefficients, and that the mirror map $z_X$ is commensurable to $J$.

\subsec{Other deformations of K3 surfaces}

We now consider three other families of K3 surfaces: quartic hypersurfaces
$X_{(4)}$ in ${\bf P}^3$, the complete intersections of quadrics and cubics
$X_{(2,3)}$ in ${\bf P}^4$ and the complete intersections of 3 quadrics
$X_{(2,2,2)}$in ${\bf P}^5$. The following table lists the types of abelian
group $G$ for the orbifold constructions, the 1-parameter deformations, the
data \potential~ for the corresponding n=2 Schwarzian equation \seA, \seBsolve.
We also include the previous example in ${\bf P}^3[1,1,1,3]$, which we denote
$X_{(6)}$.
$$
\vbox{\offinterlineskip\tabskip=0pt
\halign{\strut\vrule#
&\hfil~$#$
&\vrule#
&~~$#$~~\hfil
&~~$#$~~\hfil
&~~$#$~~\hfil
&\vrule#\cr
\noalign{\hrule}
& X,G && deformations &Q(z)~{\rm in~\seA }& & \cr
\noalign{\hrule}
&X_{(6)},(6,6,2) && x_1^6+x_2^6+x_3^6+x_4^2+z^{-1/6}x_1x_2x_3x_4=0&
\displaystyle{1 - 1968 z + 2654208 z^2\over4 z^2  (-1 + 1728 z)^2}&&\cr
&X_{(4)},(4,4)&& x_1^4+x_2^4+x_3^4+x_4^4+z^{-1/4}x_1x_2x_3x_4=0&
\displaystyle{1 - 304 z + 61440 z^2\over4 z^2  (-1 + 256 z)^2}&&\cr
&X_{(2,3)},(2,2,3) && x_1^2+x_2^2+x_3^2+z^{-1/5}x_4x_5=0&
\displaystyle{1 - 132 z + 11340 z^2\over4 z^2  (-1 + 108 z)^2}&&\cr
& && x_4^3+x_5^3+z^{-1/5}x_1x_2x_3=0& &&\cr
&X_{(2,2,2)},(2,2,2) && x_1^2+x_2^2+z^{-1/6}x_3x_4=0&
\displaystyle{1 - 80 z + 4096 z^2\over4 z^2  (-1 + 64 z)^2}&&\cr
& && x_3^2+x_4^2+z^{-1/6}x_5x_6=0& &&\cr
& && x_5^2+x_6^2+z^{-1/6}x_1x_2=0& &&\cr
\noalign{\hrule}}
\hrule}$$
The respective values of $(r_3,r_1)$ in eqn. \pfB~ are $(1728,1104)$,
$(256,176)$, $(108,78)$, $(64,48)$.
The corresponding fundamental solutions in the 4 cases have Fourier series:
\eqn\kthreeqseries{\eqalign{
z_{X_{(6)}}(q)&=q - 744 q^2  + 356652 q^3  - 140361152 q^4  + 49336682190 q^5
\cr&-16114625669088 q^6+...\cr
z_{X_{(4)}}(q)&= q - 104 q^2 + 6444 q^3 - 311744 q^4 + 13018830 q^5 - 493025760
q^6+...\cr
z_{X_{(2,3)}}(q)&=q - 42 q^2  + 981 q^3  - 16988 q^4  + 244230 q^5  - 3089394
q^6  +...\cr
z_{X_{(2,2,2)}}(q)&= q - 24 q^2  + 300 q^3  - 2624 q^4  + 18126 q^5  - 105504
q^6  +...
}}

We claim that, as for $z_{X_{(6)}}$, all the other mirror maps
are commensurable with the J-function (or equivalently with $z_W$). By
numerical experiment, we identified the following modular relations:
\eqn\polyrel{\eqalign{
P_{X_{(4)}}&(z_W,z_{X_{(4)}}):=-z_W^2  + z_W z_{X_{(4)}} - 432 z_W^2
z_{X_{(4)}} - 207 z_W z_{X_{(4)}}^2  - 62208 z_W^2  z_{X_{(4)}}^2  \cr&-
z_{X_{(4)}}^3
  	+ 3456 z_W z_{X_{(4)}}^3  - 2985984 z_W^2  z_{X_{(4)}}^3 = 0\cr
P_{X_{(2,3)}}&(z_W, z_{X_{(2,3)}}):= z_W^2  - z_W z_{X_{(2,3)}} + 576 z_W^2
z_{X_{(2,3)}} + 126 z_W z_{X_{(2,3)}}^2  \cr &+
	110592 z_W^2  z_{X_{(2,3)}}^2
- 2944 z_W z_{X_{(2,3)}}^3  + 7077888 z_W^2  z_{X_{(2,3)}}^3  +
z_{X_{(2,3)}}^4=0\cr
P_{X_{(2,2,2)}}&(z_W, z_{X_{(2,2,2)}}):=
z_W^2  - z_W z_{X_{(2,2,2)}} + 624 z_W^2  z_{X_{(2,2,2)}} +
96 z_W z_{X_{(2,2,2)}}^2\cr&  + 129840 z_W^2  z_{X_{(2,2,2)}}^2  -
2352 z_W z_{X_{(2,2,2)}}^3  + 9018880 z_W^2  z_{X_{(2,2,2)}}^3 \cr& +
10495 z_W z_{X_{(2,2,2)}}^4  + 2077440 z_W^2  z_{X_{(2,2,2)}}^4  +
z_{X_{(2,2,2)}}^5 \cr& - 1488 z_W z_{X_{(2,2,2)}}^5  + 159744 z_W^2
z_{X_{(2,2,2)}}^5 + 4096 z_W^2  z_{X_{(2,2,2)}}^6=0.
}}
We now proceed to prove the first of the relations \polyrel. The other cases
are similar. By solving $P_{X_{(4)}}(w,z_{X_{(4)}}(q))=0$ for $w$, we see that
there is a branch of $w$ which admits a q-series expansion near $q=0$ and which
has the leading behavior $w(q)\sim q$.
Now it is enough to show that every branch of $w$ solves the Schwarzian
equation \seA~ with \potentialW. For then $w(q)$ above must coincide with the
fundamental solution, by uniqueness. Now applying the relation
$P_{X_{(4)}}(w(q) ,z_{X_{(4)}}(q))=0$, we can compute $w'(q),w''(q),w'''(q)$ in
terms of $w(q)$, $z_{X_{(4)}}(q)$ and its derivatives. Substituting these
expressions into \seA~ and using the fact that $z_{X_{(4)}}(q)$ satisfies its
Schwarzian equation, we find that \seA~ with \potentialW~ holds identically.

%
%
\newsec{Uniform proof of integrality}

Applying the modular relations derived above between the mirror maps and the
j-function, we will now prove that the Fourier coefficients of the mirror maps
are integral. We begin with the following lemma. Let $z_0(q)$ be an integral
q-series and
\eqn\dumb{P(x,y)=\sum a_{i,j} x^iy^j}
be a nonzero polynomial with integral coefficients $a_{i,j}$. Let
$z(q)=\sum_{n\geq1}c_nq^n$ with $c_1=1$. Let $l:=min\{i+j|a_{i,j}\neq0\}$. Now
suppose that
\eqn\mvalue{m:=\sum_{i\geq0}a_{l-i,i}i\neq0.}
We claim that the relation $P(z_0(q),z(q))=0$ determines $z(q)$ uniquely and
that the coefficients $c_n$ are in ${\bf Z}[{1\over m}]$.

Consider the coefficient $K_N$ of $q^N$, for $N>l$, in the series
$P(z_0(q),z(q))$. Since $z_0(q)^j=q^j+O(q^{j+1})$ for $j\geq0$, the
contribution to $K_N$ from the term
$a_{p-i,i}z_0(q)^{p-i} z(q)^i$ ($p\geq l$) must be an integral linear
combination of monomials of the form $c_{n_1}\cdots c_{n_i}$, with $n_1+\cdots+
n_i\leq N-(p-i)$. From this inequality, we find that $n_k\leq N-p+i-(n_1+\cdots
n_i-n_k)\leq N-p+1$ for $k=1,..,i$. Thus the contribution to $K_N$ from the
term
$a_{p-i,i}z_0(q)^{p-i} z(q)^i$ lives in ${\bf Z}[c_1,..,c_{N-p+1}]$. In
particular, for $p>l$ this contribution is in ${\bf Z}[c_1,..,c_{N-l}]$.

Consider the case $p=l$. It is easy to see that the order $q^N$ term in
$a_{l-i,i}z_0(q)^{l-i} z(q)^i$ takes the form
\eqn\dumb{a_{l-i,i} q^{l-i} i q^{i-1}c_{N-l+1} q^{N-l+1} + g q^N}
where $g\in {\bf Z}[c_1,..,c_{N-l}]$. Setting $c_1=1$, we see that
\eqn\dumb{K_N= m c_{N-l+1} + h}
for some $h\in {\bf Z}[c_1,..,c_{N-l}]$. By induction, the relations $K_N=0$
imply that $c_{N-l+1}\in {\bf Z}[{1\over m}]$.

It follows immediately from the above lemma that if $m=\pm 1$, then $z(q)$ is
an  integral q-series.

Now consider the polynomial relation between $z_W(q)$ and the mirror map $z(q)$
for each case of elliptic curves (see table in section \elliptic). For each
of the polynomials, we have $l=1$, $m=1$. It follows that $z(q)$ is integral
in each case. Now for the four cases of
K3 surfaces, we found the relations \polyrel. For these polynomials, we have
$l=2$, $m=1$. It follows again that $z(q)$ is integral in each case.

We should remark that the above lemma gives us an effective way to construct
many integral q-series $f(q)$ which are commensurable with $1/1728J(q)$. For
example let $r(x,y)$ be any integral polynomial with no constant or linear
terms. Then our lemma implies that the equation
\eqn\dumb{{1\over 1728J(q)} -z(q) +r({1\over 1728J(q)},z(q)) =0.}
determines a unique integral q-series $z(q)$. This also implies that $z(q)$
admits an analytic continuation (with at most finitely many sheets) to the
upper half plane $Im\ t>0$. Moreover, since $J(q)$ satisfy the Schwarzian
equation $\seA$ with rational $Q(z)$, $z(q)$ also satisfies a similar equation
with algebraic $Q(z)$.

\subsec{Relations with genus zero functions and the Thompson series}

In this section, we point out some tantalizing evidence that the mirror maps
for K3 surfaces are possibly related to the theory of the Griess-Fischer group.
We will in fact give an alternative proof of integrality for the mirror maps.
This proof, though more technical than the previous one, suggests some deeper
connection with the theory of modular functions and the Thompson series.

Let's us recall a few facts about genus zero functions and the Thompson series
(see \ref\CN{J.H. Conway and S.P. Norton,``Monstrous Moonshine'', Bull. London
Math. Soc., 11(1979) 308-339.}.)  Let $H$ be the upper half plane, $G$ be a
discrete subgroup of $PSL(2,{\bf R})$ which acts on $H$ by linear fractional
transformations.  We consider  meromorphic functions $f$ on $H$ which has a
Fourier expansion $\sum_{n>m}a_n q^n$ for $Im\ t>>0$ ($q=e^{2\pi i  t}$).

If $f$ is invariant under $G$, then $f$ defines a function on the quotient
$H/G$. If $H/G$ is a genus zero Riemann surface with finitely many punctures,
we call $G$ a genus zero group and $f$ a genus zero function.
In this case,  the  field of genus zero functions for $G$ has a canonical
generator $h$ known as the normalized hauptmodul for $G$. It has a Fourier
expansion of the form $h(t)={1\over q}+O(q)$. For example
$\Gamma_0(1):=PSL(2,{\bf Z})$ is a genus zero group and the Dedekind-Klein
j-function $j(t)-744$ is the hauptmodul for this group.  More generally, a list
of 174 hauptmoduls corresponding to certain congruent subgroups is known
\ref\antwerpIV{B.J. Birch and W. Kuyk, ed. {\sl Modular Functions of One
Variable IV}, Proc. Intern. Summer School (Antwerp, 1972, Springer).}.  Let
$\Gamma_0(N)$ be the group consisting of integral transformations $t\rightarrow
(at+b)/(ct+d)$ with $N|c$. Then the 174 genus zero groups are subgroups of
$PSL(2,{\bf R})$ generated by $\Gamma_0(N)$ together with a set of Atkin-Lehner
involutions  (see \ref\antwerpV{P.G. Kluit,``On the Normalizer of
$\Gamma_0(n)$'', in {\sl Modular Functions of One Variable V} (Bonn, 1976,
Springer), 239-246.}).

It turns out that there is a deep connection between the theory of modular
functions and the Griess-Fischer group $M$ also known as the Monster. It is the
largest sporadic simple finite group, whose order is:
\eqn\dumb{2^{46}.3^{20}.5^9.7^6.11^2.13^3.17.19.23.29.31.41.59.71.}
 For a long and glory history of this group, see the paper of Conway-Norton
\CN.  Based on some remarkable observations of MacKay and Thompson,
Conway-Norton conjectured that there exist a natural ${\bf Z}$-graded
representation $V$ of $M$ with the following property: for every $g\in M$, the
q-series, called the Thompson series of $g$,
\eqn\thompson{T_g(q):=\sum tr(g|_{V_n}) q^{n-1}}
is a normalized hauptmodul in the list of 174 mentioned above. It turns out
that all but 3 of those 174 hauptmoduls correspond to Thompson series. At the
time the conjecture was made, neither $M$ nor $V$ had been known to exist,
though the evidence for them was overwhelming. On some hypotheses, the first
few coefficients of the Thompson series were computed in \CN~ and were seen to
coincide with the coefficients of the appropriate hauptmoduls. (see table 4 in
\CN).
Later, Griess proved that $M$ exists by constructing a 196883-dimensional
algebra $B$ in which $M$ acts by automorphisms. Subsequently,
Frenkel-Lepowsky-Meurman \ref\flm{I. Frenkel, J. Lepowsky and A. Meurman, {\sl
Vertex Operator algebras and the monster}, Academic Press, Boston 1988.}
constructed a ${\bf Z}$-graded representation $V^\#$ of $M$ with the property
that $V^\#_2$ is the direct sum of $B$ with the 1-dimensional representation of
$M$ and that $\sum tr(1|_{V^\#_n}) q^{n-1}=j(q)-744$. Borcherds \ref\bor{R.
Borcherds, ``Monstrous moonshine and monstrous Lie superalgebras'', Invent.
Math. 109 (1992) 405-444.} finally completed the proof of the Conway-Norton
conjecture by showing that $V^\#$ does indeed have the property that the
corresponding Thompson series coincide with the hauptmoduls suggested in \CN.

Our brief account of the Monster by no means does justice to the many of the
remarkable developments surrounding the theory of the Monster. The above
paragraph is meant to provide just enough background to enter a discussion on
the relation, which we are about to show, between the Thompson series and our
mirror maps. Based on numerical experiments, we have

{\it Observation: The reciprocal of the four mirror maps \kthreeqseries~ agree
respectively, up to an additive constant, with the Thompson series
$T_{1A},T_{2A}, T_{3A}, T_{4A}$ (see table 4 of \CN).}

Before we give a proof of this assertion, we argue that together with the
Conway-Norton conjecture, the above assertion implies the integrality of mirror
maps \kthreeqseries. The character values of a finite group are algebraic
integers. But since $V^\#$ is defined over the rationals,
the coefficients of the Thompson series which are character values of $M$, must
be rational integers. Note also that this property can also be deduced from the
explicit representation of the hauptmoduls in terms of the Dedekind eta
function.

To prove our assertion above, we will use Schwarz' theorem on the Riemann
mapping and the fact that the Thompson series above are hauptmoduls for the
following genus zero groups $\Gamma_0(1)$, $\Gamma_0(2)+$, $\Gamma_0(3)+$,
$\Gamma_0(4)+$ (see \CN~ on notations). If $G$ is a genus zero group and $h(t)$
its normalized hauptmodul, then it is easy to check that the expression
$\{h(t),t\}/(2h'(t)^2)$ is a meromorphic function which is invariant under $G$.
 Since $h(t)$ is a generator of the function field for $G$, it follows that
$\{h(t),t\}/(2h'(t)^2)$ is a rational expression $Q$ in
$h(t)$. This shows that a hauptmodul satisfies a Schwarzian equation of the
type \seA. Our problem is to construct $Q$ in each of the cases of interest.
Once we determine the Schwarzian equations governing the hauptmoduls
$T_{1A},T_{2A}, T_{3A}, T_{3A}$ in each case, it is enough to see that they
coincide with the
corresponding Schwarzian equation which we already know governing  ${1\over
z}+c$ where $z$ is one of the $z_{X_{(6)}}, z_{X_{(4)}}, z_{X_{(2,3)}},
z_{X_{(2,2,2)}}$. For if two q-series satisfy the same Schwarzian equation and
has the same asymptotic behavior  $~1/q$, then they must coincide by
uniqueness.
The constant $c$ above will turn out to be the coefficient of $q^2$ in each
case (so that ${1\over z}+c$  has no constant term.)

We now proceed to determine $Q$ for each of the four hauptmoduls $h(t)$ above.
Since $h(t)$ is an isomorphism from $H/G$ onto a punctured Riemann sphere, it
defines  a univalent mapping from a simply connected region  $R$ (half a
fundamental region of $G$) onto the upper half $h$-plane. If $R$ is a polygon
with circular edges, then Schwarz' Theorem tells us that $h(t)$ satisfies the
eqn. \seA~ with $Q$ having the form
\eqn\dumb{Q(z)=\sum_{i=1}^n ({A_i\over (z-a_i)^2}+{B_i\over  z-a_i})}
where $a_1,...,a_n$ are images of the vertices of $R$ and the $A_i,B_i$ are
constants. To determine $Q$ completely it is enough to know the number $n$ of
vertices (actually a reasonable bound would suffice).  For then we can use the
first few Fourier coefficients of $h(t)$ to determine the $A_i,B_i$. To find
the number of vertices, we need to determine a domain $R$ and see that it is a
circular polygon in each case. But since we
know $G$ explicitly as a congruent subgroup in each case, constructing $R$ is
easy. We find that the $R$ corresponding to the hauptmodul $T_{kA}$
($k=1,2,3,4$) is $R=\{t\in H\ |\ 0< Im\ t<{1\over 2},|t|>{1\over \sqrt{k}}\}$.
Each of these domains is a circular polygon with 3 vertices. Using the first 6
Fourier coefficients (actually 3 is enough because the $A_i,B_i$ satisfies 3
relations.) of the $T_{kA}$, we find that the respective $Q$ for their
Schwarzian equations are:
\eqn\Qs{\eqalign{
T_{1A}:& Q(z)={19896824 + 53527 z + 36 z^2\over 144 (743 + z)^2  (744 + z)^2}
\cr
T_{2A}:& Q(z)={40640 - 96 z + z^2\over 4 (-152 + z)^2  (104 + z)^2 }\cr
T_{3A}:& Q(z)={7560 - 48 z + z^2\over 4 (-66 + z)^2  (42 + z)^2}\cr
T_{4A}:& Q(z)={2752 - 32 z + z^2\over 4 (-40 + z)^2  (24 + z)^2} }}
Now consider the four Fourier series ${1\over z}+c$ mentioned above. We know
that each one satisfies a Schwarzian equation which can be easily obtained from
that of $z$. We find that the $Q$ indeed agrees with \Qs~ in each case. This
completes the proof of our assertion.

Our observation above raises various tantalizing questions. Why do the mirror
maps of our K3 surfaces correspond to the Thompson series of the Monster group?
 What happens to the other realizations of K3 surfaces in weighted projective
spaces? Do they realize other Thompson series? On the last two questions our
joint work with A. Klemm in this direction is now underway.
Early indications have shown that the correspondence continue to hold. On the
first question, we offer the following speculative remarks. It is known that
there are K3 surfaces on which a subgroup of the Mathieu group $Mi_{24}$
acts by automorphisms \ref\mukai{S. Mukai, ``Finite groups of automorphisms of
K3 surfaces and the Mathieu group'', Invent. Math. 94 (1988) 183-221.}. It
happens that $Mi_{24}$ is also a subgroup of the Monster. Thus it makes sense
to speak of the Thompson series for those elements in $Mi_{24}$. It is possible
that the appearance of Thompson series as mirror maps is related to some
appropriate action of a subgroup of $Mi_{24}$. Precisely how this happens is
unclear at this point. However, enough evidence has convinced us the following:

{{\bf Conjecture:} {\it If $z(q)$ is the mirror map for a 1-parameter
deformations of an algebraic K3 surface from an orbifold construction, then for
some $c\in{\bf Z}$, ${1\over z(q)}+c$ is a Thompson series $T_g(q)$ for some
$g\in M$.}

It seems that the above connection between the mirror map and Thompson series
is peculiar to K3 surfaces. In the case of elliptic curves for instance, unlike
the K3 case, the correspondence there is only partial.  Applying the same
technique
as above, we find that all but the 3rd family of elliptic curves we studied in
section \elliptic, correspond to Thompson series. They are $T_{2B}$, $T_{3B}$,
and $T_{4C}$ respectively. The mirror map for the 3rd family is in fact a
double-valued function on the upper half plane,  hence cannot be a modular
function of any type.

\newsec{Congruences of the Quantum Coupling}

We now move on to Calabi-Yau threefolds. We consider a 1-parameter family of
hypersurfaces or complete intersections $X_z$ in ${\bf P}^n[w]$, where $z$ is
the complex structure deformation coordinate as defined in section \defcoord~
in terms of a finite orbifold group $G$. In this case, the period vector
$\omega=(\omega_0,\omega_1,\omega_2,\omega_3)$ of the holomorphic 3-form for
the mirror manifold $X^*$ satisfy a 4th order Picard-Fuchs equation which has
the form \pftemp. The prepotential is defined as %
\eqn\dumb{F:={1\over \omega_0^2}(\omega_3\omega_0+\omega_1\omega_2)}
which is a holomorphic section of a line bundle over the moduli space $M$.
It is also related to the period vector by \cdgp\ref\specialgeom{A. Strominger,
\cmp133(1990)163;  P. Candelas and X. della Ossa,\npb355(1991)455}\ref\tian{G.
Tian, in {\sl Mathematical Aspects of String
Theory}, ed. S. T. Yau, (World Scientific, Singapore 1987)}~
$\omega=\omega_0(1,t,\partial_tF,2F-t\partial_tF)$ where $t$ is a special
coordinate defined by $t=\omega_1(z)/\omega_0(z)$. Let
\eqn\dumb{K=\partial_t^3F.}
The mirror hypothesis \cdgp~ identifies $t$ with the flat coordinate on the
Kahler cone of $X$ and asserts that $K$ is the quantum coupling for $X$, which
has the form\cdgp\ref\am{P. Aspinwall and D. Morrison, Commun. Math. Phys. 151
(1993) 45.}\ref\witt{E. Witten, Commun. Math. Phys. 118 (1988) 411.}
\eqn\instanton{K=\int_X J\wedge J\wedge J+\sum_{d\geq1}{d^3n_d e^{2\pi
itd}\over 1-e^{2\pi itd}}.}
Here $n_d$ is the ``number'' of rational curves of degree $d$ in a generic
deformation of $X$, and $J$ is the Kahler class on $X$.

In this section, we would like to study some arithmetic properties of $K$ as a
$q$-series ($q=e^{2\pi it}$) under the {\it assumption of the mirror hypothesis
with the $n_d$ integers, and that the mirror map $z(q)$ has integral Fourier
coefficients}. Specifically, we will study congruences mod $p$ of the instanton
numbers under the above hypothesis, using a formal nonlinear ODE.
We should point out that the integrality of $z(q)$ has been checked numerically
in many examples of threefolds up to at least order $q^{30}$, and has been
proved for K3 surfaces and elliptic curves in numerous cases.

\subsec{Differential equations mod $p$}

In \klryA\klryB, we derived certain polynomial diferential equations for
$z(q),K(q)$. An important feature of these equations is that they are defined
over the rationals. This is so because the Picard-Fuchs equation, from which
our ODE are constructed, are defined over the rationals. The main idea here is
to derive the $mod~p$ version of our nonlinear ODE from the Picard-Fuchs
equation. We then use some data -- basically an integer $N_e$ -- which is
extracted from the Picard-Fuchs equation in each case to derive congruences
modulo certain prime powers. We will illustrate this in several well-known
examples.

We begin with the following form of the Picard-Fuchs equation \pftemp:
\eqn\reducedpf{\left({d^4\over dz^4} + q_2(z){d^2\over dz^2}+
q_2'(z){d\over dz}+q_0(z)\right)f(z)=0}
where $q_2(z),q_0(z)$ are defined in \ccrr. By a change of coordinate $z\mapsto
t$ (see \klryB~ for details) and using the list of solutions
$f_0(z)(1,t,\partial_tF,2F-t\partial_tF)$, we show that $z,K$ as functions of
$t$, satisfy a pair of coupled nonlinear ODEs: The simpler of the two is given
by
\eqn\rhozK{\lambda_z K^4=\rho_K ~~(*)}
where
\eqn\dumb{\eqalign{
\rho_K:=&{5^2\cdot7\,{{K'}^4} - 2^3\cdot5\cdot7\,K\,{{K'}^2}\,K'' +
      7^2\,{{K}^2}\,{{K''}^2} + 2\cdot5\cdot7\,{{K}^2}\,K'\,K^{(3)} -
      2\cdot5\,{{K}^3}\,K^{(4)}}\cr
\lambda_z:=&{e(z)z'^4\over\Delta(z)^4}\cr
e(z):=&10(-10r_2+3r_2-4r_4)z+ (-9r_2^2+300r_0r_4-46r_2r_4+64r_4^2)z^2\cr&
+2r_4(9 r_2^2  - 150 r_0 r_4 + 7 r_2 r_4 - 22 r_4^2 )z^3
+r_4^2  (-9 r_2^2  + 100 r_0 r_4 + 2 r_2 r_4 + 11 r_4^2 )z^4.
}}
Here $\Delta(z)=z(1-r_4 z)$ is the discriminant of the equation \pftemp.
(Actually, the following discussion requires only that $e(z),\Delta(z)\in{\bf
Z}[z]$, with $\Delta(z)$ having leading coefficient 1. No specific forms need
to be assumed.) We {\it define
$N_e$ to be the l.c.m. of the coefficients in $e(z)$.} Then it is clear that
$n|N_e$ iff $n$ divides all Fourier coefficients of $e(z(q))$. {\it All the
arithmetic properties we derive of $K(q)$ will be entirely controlled by this
one integer $N_e$ and the classical intersection number
$K_{cl}:=\int_X J\wedge J\wedge J$.}

Notice that (a) eqn. (*) is defined over ${\bf Z}$; (b) eqn. (*) is invariant
under an overall constant scaling of $K$; (c) the primes 2,5,7 feature
prominantly. It turns out that, though less obvious, the eqn. also simplifies
considerably modulo 3. The main purpose of having eqn. (*) is as follows.
Suppose a 1-parameter family of mirror Calabi-Yau varieties is given such that
$p|N_e$ for some prime $p$. Then the left hand side of (*) becomes zero over
the field of $q$-series ${\bf Z}/p{\bf Z}((q))$. We can then study the
$q$-series solution $K$ \instanton of this simplified equation modulo $p$ or
its powers. We should point out that in \klryA\klryB~, we derived a similar
equation but over ${\bf Q}$. The result there was much more complicated.

{\it Notations:} In the following discussion, $K$ will be the quantum coupling
\instanton. The $'$ in the equation (*) means $q{d\over dq}$. Thus
$(q^n)'\equiv0~mod~p$ whenever $p|n$. More generally when $p$ is prime and
$h=\sum a_nq^n$ is any integral q-series, the $pth$ derivative $h^{(p)}\equiv
h'$ because $n^p\equiv n$ for all $n$. In what follows, we will use these few
tricks repeatedly, sometimes without mention, to simplify our computations. We
write $n|h$ if $n|a_n$ for all $n$. We also write $\rho_h$ to denote the right
hand side of \rhozK~ but with $K$ replaced with $h$:
\eqn\rhodef{\eqalign{
\rho_h:=&{5^2\cdot7\,{{h'}^4} - 2^3\cdot5\cdot7\,h\,{{h'}^2}\,h'' +
      7^2\,{{h}^2}\,{{h''}^2} + 2\cdot5\cdot7\,{{h}^2}\,h'\,h^{(3)} -
      2\cdot5\,{{h}^3}\,h^{(4)}}.
}}
For each prime $p$, we introduce the notation:
\eqn\dumb{h_p:= \sum_{p\not{|}n}a_nq^n.}
Thus we have $h'\equiv h_p'~mod~p$. Finally, given $K$ \instanton, we denote
the largest integer divisor of $K$ by $m(K)$.

\subsec{mod p}
\seclab\modp

{\it Claim 1: Let $h:={K\over m(K)}$. For any integer $n$, $n|h'$ implies that
$n|\rho_h$, and $n|\rho_h$ implies that $n|N_e$. Conversely, if $n|N_e$ then
$n|\rho_h$.}

Pf: If $n|h'$ then from \rhodef, it follows that $n|\rho_h$. Assuming
$n|\rho_h$, eqn. (*) becomes $e(z) z'^4 h^4\equiv0~mod~n$. Since neither $h$
nor $z'$ has any overall factor, we have $e(z)\equiv0$. This implies that
$N_e\equiv0$. The converse follows immediately from eqn. (*).

This says that the integer $N_e$ tells us much about congruences of ${K\over
m(K)}$.

{\it Claim 2: Fix a prime $p$ and suppose $p^r|K_p$. Then $p^r|n_d$ for all
$d\not\equiv 0~mod~p$, and $p^{min(3,r)}|(K-K_{cl})$.}

Pf: From \instanton, the $mth$ ($m>0$) Fourier coefficient of $K$ is
\eqn\dumb{k_m=\sum_{d|m} d^3 n_d.}
By supposition, $p^r|k_m$ for all $m\not\equiv 0~mod~p$. Let $d\not\equiv
0~mod~p$ be the smallest such integer for which $p^r\not{|}n_d$. Now $p^r|k_d$
means that $p^r|(d^3 n_d +\sum_{d'|d,~d'<d} d'^3 n_{d'})$. Since $p\not{|}d$,
that $d'|d$ implies that $p\not{|}d'$, hence  $p^r|n_{d'}$ by the minimality of
$d$. So $p^r|(d^3 n_d +\sum_{d'|d,~d'<d} d'^3 n_{d'})$ implies that $p^r|n_d$
which is a contradiction.

In particular, we have $p^r|d^3 n_d$ for $p\not{|}d$. Trivially $p^3|d^3 n_d$
for $p|d$, and so we have $p^{min(3,r)}|d^3n_d$ for all $d$. Now it follows
immediately from \instanton~ that $p^{min(3,r)}|(K-K_{cl})$.

Now let $h$ be an integral q-series. As pointed out above, $h'\equiv
h_p'~mod~p$. It follows that $p|h_p$ implies $p|\rho_h$. On the other hand, we
have the following:

{\it Claim 3: Assume prime $p\not{=}2,5$. Suppose for any integral q-series $h$
with $p\not{|}h$, $p|\rho_h$ implies $p|h_p$. Then for all $r>0$, $p^r|\rho_h$
implies that $p^r|h_p$ and $p^r|\rho_{h-h_p}$.}

Pf: If $p|\rho_h$ then $p|h_p$, and so we can write $h_p=p f$ for some integral
series $f$. Write also $g=h-h_p$, which has only q-powers $q^n$ with $p|n$.
Thus $p|g'$, hence $p|\rho_g$. Thus our statement holds for $r=1$.

Suppose it holds up to some $r>0$. Assume $p^{r+1}|\rho_h$. By inductive
hypothesis we have at least $p^r|h_p$, and so we can write $h_p=p^r f$,
$g=h-h_p$ as before. Because $h=p^rf+g$, and $p\not{|}h$ by assumption, we have
$p\not{|}g$. The equation $\rho_h\equiv0~mod~p^{r+1}$ becomes, after some
simplifications using $p^i|g^{(i)}$,
\eqn\vanish{\rho_g-10p^rg^3f''''\equiv0~mod~p^{r+1}.}
But $g$ only has q-powers $q^n$ with $p|n$, and hence so does $\rho_g$. On the
other hand $f$ (hence $f''''$) has only $q^n$ with $p\not{|}n$. Thus \vanish~
implies that $p^{r+1}|\rho_g$ and $p^{r+1}|(10p^rg^3f'''')$. From the fact that
$p\not{=}2,5$ and that $p\not{|}g$, it follows that $p|f''''$ and hence $p|f$.
Thus $h_p=p^rf$ implies that $p^{r+1}|h_p$.

The cases $p=2,5$ are dealt with separately. We will also discuss the cases
$p=3,7$ in more details.

\subsec{mod 2}
\seclab\modtwo

{\it Claim 4: Let $h$ be integral q-series with $2\not{|}h$. Then $2|\rho_h$
implies that $2|h_2$ or $2|(h-h_2)$. If moreover $2|h_2$ then (a) $2^r|\rho_h$
implies that $2^r|\rho_{h-h_2}$, and (b) $2^{r+1}|\rho_h$ implies that
$2^r|h_2$, for all $r>0$.}

Pf: Suppose $2|\rho_h$. Then we get
\eqn\dumb{h'^4+h^2h'^2\equiv0~mod~2.}
This says that either $2|h'$ or $2|(h+h')$. But $h'\equiv h_2'\equiv
h_2~mod~2$, implying that either $2|h_2$ or $2|(h-h_2)$. This proves the first
half.

For part (a), the proof of {\it Claim 3} applies to $p=2$ without change.

For part (b), we modify the inductive argument there as follows. Suppose
$2^3|\rho_h$ and write $h_2=2f$, $g=h-h_2$ as before. The equation
$\rho_h\equiv0~mod~2^3$ becomes
\eqn\dumbtwo{\rho_g+49.2^2 g^2f''^2-10.2g^3f''''\equiv0~mod~2^3.}
Part (a) implies that $2^3|\rho_g$. Also $g^2f''^2$ only has q-powers $q^n$
with $2|n$ while $g^3f''''$ has only those with $2\not{|}n$. Thus \dumbtwo
implies that all three terms vanish separately. It follows that $2|f''''$ and
hence $2|f$. Thus $h_2=2f$ implies that $2^2|h_2$.
Suppose (b) holds up to some $r>1$. Assume $2^{r+2}|\rho_h$. By inductive
hypothesis we have at least $2^r|h_2$, and so we can write $h_2=2^r f$,
$g=h-h_2$ as before. The equation $\rho_h\equiv0~mod~2^{r+2}$ becomes
\eqn\dumb{\rho_g-10.2^rg^3f''''\equiv0~mod~2^{r+2}.}
Part (a) implies that $2^{r+2}|\rho_g$.  By assumptions, $2\not{|}h$ and
$2|h_2$ hence $2\not{|}g$. It follows that $2|f''''$, hence $2|f$. Thus
$h_2=2^rf$ implies that $2^{r+1}|h_2$. This completes our proof of part (b).

{\it Claim 5:} If $2|(K+K')$ then $2|n_d$ for all odd $d$, and $2|K$.

Pf: Let $l$ be the smallest odd integer such that $2\not{|}n_l$. Now $2|(K+K')$
means that $2|k_m$ for all even $m$, where $k_m$ is the $mth$ Fourier
coefficients of $K$. In particular $2|k_{2l}$. But
\eqn\dumb{k_{2l}=\sum_{d|2l}d^3n_d=\sum_{d|2l,~d~odd}d^3n_d
+\sum_{d|2l,~d~even}d^3n_d.}
The even part obviously divides $2$, while the odd part is
\eqn\dumb{\sum_{d|2l,~d~odd}d^3n_d=l^3n_l+\sum_{d|l,~d~odd,~l>d}d^3n_d.}
By the minimality of $l$, the second sum divides $2$, implying that $2|l^3n_l$.
But $l$ is odd, hence $2|n_l$ which is a contradiction. Thus $2|n_d$ for odd
$d$.

Now consider \instanton. If $2|n_d$ for odd $d$, then obviously $2|d^3n_d$ for
all $d$. Also $2|(K+K')$ implies that $2|K_{cl}$.
So we have $2|K$.

{\it Remarks:} Let $h:={K\over m(K)}$ and suppose $2^r|N_e$, $r>0$. By {\it
Claim 1}, we have $2^r|\rho_h$. By {\it Claim 4}, we have either $2|h_2$ or
$2|(h-h_2)$. Suppose first $2|h_2$. Then $2^{r-1}|h_2$ by {\it Claim 4b}, hence
$(m(K)2^{r-1})|K_2$. Thus by {\it Claim 2}, we see that $2^{s+r-1}|n_d$ for all
$d$ odd, where $s$ is the largest integer such that $2^s|m(K)$. Moreover,
$2^{min(3,s+r-1)}|(K-K_{cl})$. To summarize: {\it if $2^r|N_e$ and $2|{K_2\over
m(K)}$, then (a) $2^{s+r-1}|n_d$ for all $d$ odd, where $s$ is the largest
integer such that $2^s|m(K)$, and (b) $2^{min(3,s+r-1)}|(K-K_{cl})$. }

Suppose now $2|(h-h_2)$ instead. This means in particular that $2|(K+K')$, and
so by {\it Claim 5} we have $2|n_d$ for all odd $d$, and $2|K$. Thus
$m(K)\geq2$.
So {\it if $2|N_e$ and $2|({K-K_2\over m(K)})$, then $2|n_d$ for all odd $d$
and $2|K$.}

\subsec{mod 3}
\seclab\modthree

Suppose $h$ is any integral series with $3|\rho_h$ and $3\not{|}h$. We get by
differentiating $\rho_h\equiv 0~mod~3$ once,
\eqn\dumb{hh' (2 h^2  + h'^2  + hh'')\equiv0~mod~3.}
If $3\not{|}h'$ then $3|(2 h^2  + h'^2  + hh'')$. Differentiating $2 h^2  +
h'^2  + hh''$ and simplifying it by $h'''\equiv h'~mod~3$, we get $3|(2hh')$
implying that $3|h'$, hence $3|h_3$. Thus we have shown that $3|\rho_h$ implies
$3|h'$ and $3|h_3$.

{\it Remark:} Let $h:={K\over m(K)}$ and suppose $3^r|N_e$, $r>0$. Then by {\it
Claim 1}, we have $3^r|\rho_h$. We have just shown that $3|\rho_h$ implies
$3|h_3$. Then $3^r|h_3$ by {\it Claim 3}, hence $(m(K)3^{r})|K_3$. Thus by {\it
Claim 2}, we see that $3^{s+r}|n_d$ for all $d\not\equiv 0~mod~3$, where $s$ is
the largest integer such that $2^{s}|m(K)$. Moreover,
$3^{min(3,s+r)}|(K-K_{cl})$. To summarize: {\it if $3^r|N_e$, then (a)
$3^{s+r}|n_d$ for all $d\not\equiv 0~mod~3$, where $s$ is the largest integer
such that $3^{s}|m(K)$, and (b) $3^{min(3,s+r)}|(K-K_{cl})$.}

\subsec{mod 5}
\seclab\modfive

{\it Claim 6: Let $h$ be integral q-series with $5\not{|}h$. Then $5|\rho_h$
implies that $5|h_5$. Moreover we have (a) $5^r|\rho_h$ implies that
$5^r|\rho_{h-h_5}$, and (b) $5^{r+1}|\rho_h$ implies that $5^r|h_5$, for all
$r>0$.}

Pf: Suppose $5|\rho_h$. Then the eqn. $\rho_h\equiv 0~mod~5$ reads
$4hh''\equiv0~mod~5$. Hence $5|h''$ which implies $5|h_5$. This proves the
first statement.

For part (a), the proof of {\it Claim 3} applies to $p=5$ without change.

For part (b), the proof of {\it Claim 4b} applies here with only minor changes.

{\it Remarks:} Let $h:={K\over m(K)}$ and suppose $5^r|N_e$, $r>0$. By {\it
Claim 1}, we have $5^r|\rho_h$. Then $5^{r-1}|h_5$ by {\it Claim 6b}, hence
$(m(K)5^{r-1})|K_5$. Thus by {\it Claim 2}, we see that $5^{s+r-1}|n_d$ for all
$d\not\equiv 0~mod~5$, where $s$ is the largest integer such that $5^{s}|m(K)$.
Moreover, $5^{min(3,s+r-1)}|(K-K_{cl})$. To summarize: {\it if $5^r|N_e$, then
(a) $5^{s+r-1}|n_d$ for all $d\not\equiv 0~mod~5$, where $s$ is the largest
integer such that $5^{s}|m(K)$, and (b) $5^{min(3,s+r-1)}|(K-K_{cl})$.}

\subsec{mod 7}
\seclab\modseven

We summarize the result here: {\it if $7^r|N_e$, then (a) $7^{s+r}|n_d$ for all
$d\not\equiv 0~mod~7$, where $s$ is the largest integer such that $7^{s}|m(K)$,
and (b) $7^{min(3,s+r)}|(K-K_{cl})$.} The argument here is quite similar to the
case of $mod~3$.

\subsec{Examples}

Under the assumption of the integrality of the mirror map $z(q)$ and the mirror
hypothesis' assertion that quantum coupling $K(q)$ is given by \instanton with
integers $n_d$, we consider the following examples applying the above results:

1. Let $X$ be the Fermat quintics in ${\bf P}^4$. The orbifold group here is an
abelian group of type $(5,5,5)$\gp\cdgp. In this case, we have $N_e=-5^3$ and
$K_{cl}=5$. It follows from our $mod~5$ analysis that at least
$5^2|(K-K_{cl})$. But since $K_{cl}=5$, we have $5|K$ and $m(K)=5$. Thus
$5^3|n_d$ for all $d\not{\equiv}0~mod~5$. Because $n_1=23.5^3$, $5^3$ is the
upper bound on the prime power dividing all $n_d$. Experimentally however, it
seems that $5^3$ divides {\it all} $n_d$, as previously observed by others (see
table 4 of \cdgp). Note that since $N_e$ has no prime divisor other than $5$,
there is no other prime $p$ dividing all $n_d$, by {\it Claim 1}.

2. Consider $X$ the complete intersection of a quartic $x_1^4+x_2^4+x_3^4=0$
and a sextic $x_4^3+x_5^3+x_6^2=0$ in ${\bf P}^5[1,1,1,2,2,3]$. The orbifold
group is of type $(4,4,3)$. In this case, we have $N_e=-2^6.3^5$ and
$K_{cl}=2$.
 It follows from our $mod~3$ analysis that $3^5|n_d$ for all
 $d\not{\equiv}0~mod~3$. Once again, this lower bound on the powers of $3$ is
sharp because already $n_1=2^6.3^5$. For $mod~2$, our analysis shows that
either $2|h_2$ our $2|(h-h_2)$, where $h={K\over m(K)}$. Suppose $2|(h-h_2)$.
Then our analysis above implies that $2|K$ and $m(K)\geq2$. But $K_{cl}=2\geq
m(K)$, implying that $m(K)=2$. Thus $(h-h_2)$ has constant term ${K_{cl}\over
m(K)}=1$, contradicting $2|(h-h_2)$. It follows that we must have $2|h_2$. By
the $mod~2$ analysis above, we conclude that at least $2^6|n_d$ for all $d$
odd. Again this lower bound on the power of $2$ is also sharp because already
$n_1=2^6.3^5$.  Note that since $N_e$ has no prime divisor other than $2,3$,
there is no other prime $p$ dividing all $n_d$, by {\it Claim 1}.

3. Finally consider $X$ the degree 10 hypersurface of Fermat type in ${\bf
P}^4[1,1,1,2,5]$. The orbifold group is of type $(10,5,2)$. In this case, we
have $N_e=-2^6.5^3$ and $K_{cl}=1$. Thus $m(K)=1$. It follows from our $mod~2$
and $mod~5$ analyses that at least $2^5|n_d$ for $d$ odd; and $5^2|n_d$ for
$d\not{\equiv}0~mod~5$. These bounds on the prime power are also sharp because
already $n_1=2^5.5^2.17$. Note that since $N_e$ has no prime divisor other than
$2,5$, there is no other prime $p$ dividing all $n_d$, by {\it Claim 1}.

\listrefs
\end